\documentclass[twocolumn,superscriptaddress,amssymb, nobibnotes, aps, prb]{revtex4-2}
\usepackage{amsmath}
\usepackage{graphicx}
\usepackage{amsfonts}
\usepackage{amssymb}
\usepackage{bm}
\usepackage{dsfont}
\usepackage{epsfig,float,afterpage}
\usepackage[colorlinks,allcolors=blue]{hyperref}
\usepackage[usenames,dvipsnames]{color}

\newcommand{\beq}{\begin{equation}}
\newcommand{\eeq}{\end{equation}}
\newcommand{\bes}{\begin{subequations}}
\newcommand{\ees}{\end{subequations}}
\newcommand{\bea}{\begin{eqnarray}}
\newcommand{\eea}{\end{eqnarray}}
\newcommand{\ba}{\begin{array}}
\newcommand{\ea}{\end{array}}
\newcommand{\beqn}{\begin{eqnarray*}}
\newcommand{\eeqn}{\end{eqnarray*}}

\newcommand{\nn}{\nonumber}

\newcommand{\tl}[1]{\tilde{#1}}
\newcommand{\bra}{\langle}
\newcommand{\ket}{\rangle}

\def\nn{\nonumber}

\begin{document}

\title{Heat transport through an open coupled scalar field theory hosting stability-to-instability transition}
\author{T. R. Vishnu and Dibyendu Roy}
\affiliation{Raman Research Institute, Bangalore 560080, India}

\begin{abstract}

We investigate heat transport through a one-dimensional open coupled scalar field theory, depicted as a network of harmonic oscillators connected to thermal baths at the boundaries. The non-Hermitian dynamical matrix of the network undergoes a stability-to-instability transition at the exceptional points as the coupling strength between the scalar fields increases. The open network in the unstable regime,  marked by the emergence of inverted oscillator modes, does not acquire a steady state, and the heat conduction is then unbounded for general bath couplings. In this work, we engineer a unique bath coupling where a single bath is connected to two fields at each edge with the same strength. This configuration leads to a finite steady-state heat conduction in the network, even in the unstable regime. We also study general bath couplings, e.g., connecting two fields to two separate baths at each boundary, which shows an exciting signature of approaching the unstable regime for massive fields. We derive analytical expressions for high-temperature classical heat current through the network for different bath couplings at the edges and compare them. Furthermore, we determine the temperature dependence of low-temperature quantum heat current in different cases. Our study will help to probe topological phases and phase transitions in various quadratic Hermitian bosonic models whose dynamical matrices resemble non-Hermitian Hamiltonians, hosting exciting topological phases.

\end{abstract}

\begin{titlepage}
\maketitle
\end{titlepage}

\section{Introduction}
\label{s:Introduction}

The physics of non-Hermitian quantum systems \cite{hatano_localization_1996} has received extensive research interest in the last three decades. The first wave of research came through the parity-time $(\mathcal{PT})$ symmetric non-Hermitian Hamiltonians, which display real eigenenergies and exhibit exciting comparisons to the dynamics of Hermitian Hamiltonians \cite{Bender1998, KHARE200053, BAGCHI200079, Mostafazadeh_2002, Bender_2002_generalized, berry_physics_2004, Guo2009, heiss_physics_2012}. The topological features of non-Hermitian Hamiltonians with different discrete symmetries \cite{rudner_topological_2009, liang_topological_2013, shen_topological_2018, lieu_topological_2018, pan_photonic_2018, kawabata_symmetry_2019, wang_topological_2021, vyas_topological_2021, ritutopo2022} have generated a second wave of research interest. While the Hermitian Hamiltonian description is valid for an isolated physical system, the non-Hermitian modeling is natural for most systems when they are in contact with one or multiple environments, as in the canonical and grand-canonical descriptions of equilibrium statistical mechanics. There is another interesting class of Hermitian quadratic bosonic Hamiltonians \cite{Colpa1978, Rossignoli2005} for closed systems, whose dynamics are governed by non-Hermitian Bogoliubov-de Gennes effective Hamiltonians \cite{McDonald_2018, Lieu2018, Wang2019, Flynn_2020, Flynn_2021}. The emergent non-Hermitian effective Hamiltonians can have a generalized $\mathcal{PT}$ symmetry, which is broken when diagonalisability is lost at exceptional points (EPs) in the parameter space. These EPs mark stability-to-instability transitions between a dynamically stable regime with the unbroken $\mathcal{PT}$ symmetry and a dynamically unstable regime with the broken $\mathcal{PT}$ symmetry.

We consider a coupled quantum scalar field theory in one space dimension that exemplifies such a Hermitian quadratic bosonic Hamiltonian. Such field theories have been explored for decades in understanding basic physical phenomena, such as the Higgs field, and introducing novel concepts and techniques of quantum field theories \cite{Peskin1995, Schwartz2014}. It is possible to engineer prototypes of such field theories using mechanical, optical, and opto-mechanical networks \cite{Aspelmeyer2014}. In the unstable regime of the coupled scalar field theory or network, it hosts inverted harmonic oscillator \cite{Barton1986} modes that have been examined for a diverse set of physical phenomena, including the Hawking-Unruh effect and scattering in the lowest Landau level in quantum Hall systems \cite{Varsha2021}. Here, we connect the two boundaries of the coupled scalar field theory discretized on a lattice as an oscillator network to heat baths kept at different temperatures. We investigate heat transport \cite{Lepri2003,Dhar2008} in this open network. Our motivation is to understand how heat transport behaves around the EPs marking stability-to-instability transitions \cite{Xu2023}. To study heat transport, we write quantum Langevin equations for the degrees of freedom of the scalar field theory after integrating out the baths, and solve these equations using the non-equilibrium Green's functions \cite{Dhar_2006, Dhar2008}. The steady-state heat transport emerges only in the dynamically stable regime of the field theory for general system-bath coupling due to the appearance of unbounded inverted oscillator modes in the unstable regime. Nevertheless, as explained in our study, heat transport is insensitive to stability-to-instability transition for some particular types of bath couplings, which is one of our main findings. We derive analytical expressions for the heat current in the high-temperature linear response regime and evaluate the temperature dependence of quantum heat current. The temperature dependence is different for the massive and massless scalar fields and it also depends on the bath coupling and the coupling between the scalar fields. We further compare steady-state heat transport between different types of bath couplings at two boundaries and between separate heat bath spectral properties. Heat current for general bath couplings, e.g., connecting two fields to two separate baths at each boundary, shows an exciting signature of approaching the unstable regime for massive scalar fields.  

The rest of the paper is divided into six sections and three appendices. We introduce the non-Hermitian description for the dynamical matrix of the coupled scalar field theory and the stability-to-instability transition in Sec.~\ref{s:Dynamical-matrix}. In Sec.~\ref{s:Lattice}, we introduce a lattice version of the field theory and its spectral properties. The heat conduction through the lattice of oscillators, connected differently to heat baths at the boundaries is discussed in Sec.~\ref{s:Heat}. We include our results for single and two baths at each edge in Sec.~\ref{s:Single-bath} and Sec.~\ref{s:Two-baths}, respectively. We conclude with a summary and outlook in Sec.~\ref{s:Summary}. Details of our derivation of heat current for two different bath couplings and different bath spectral properties are included in the three appendices at the end.

\section{Dynamical matrix and stability-to-instability transition}
\label{s:Dynamical-matrix}

We consider a coupled quantum scalar field theory of two real scalar fields $\phi_1(x)$ and $\phi_2(x)$ with a minimal coupling represented by the Hamiltonian:
	\bea
	\mathcal{H} &=& \frac{1}{2} \int dx\: \Big[ \Pi_1^2 + c^2(\partial_x \phi_1)^2 + \Pi_2^2 + c^2(\partial_x \phi_2)^2  \cr 
	&+& \frac{m^2c^4}{\hbar^2} (\phi_1^2 + \phi_2^2) \Big] + \lambda c^2\int dx\: (\partial_x \phi_1)(\partial_x \phi_2),
	\label{e:Coupled-FT-Hamiltonian}
	\eea
        where the field variables satisfy the usual canonical commutation relations, $[\phi_{\alpha}(x),\Pi_{\beta}(x')] = i\hbar \delta_{\alpha \beta} \delta(x-x')$ for $\alpha, \beta = 1,2$. Here, $\lambda >0$ is a dimensionless coupling constant and $m$ is the mass of the fields. We write down the Heisenberg equations for $\phi_{\alpha},\Pi_{\alpha}$, and then take Fourier transformation to momentum space as $\tilde{\phi}_{\alpha}(q,t)= (1/\sqrt{2\pi})\int dx\, e^{-iqx/\hbar} \phi_{\alpha}(x,t)$ and $\tilde{\Pi}_{\alpha}(q,t)= (1/\sqrt{2\pi})\int dx\, e^{-iqx/\hbar} \Pi_{\alpha}(x,t)$ for $\alpha=1,2$ to find
	\bea
	i \frac{\partial \hat{\phi}}{\partial t} = i \begin{bmatrix}
	0 & 1 & 0 & 0 \\
	\frac{-m^2 c^4 - q^2 c^2}{\hbar^2} & 0 & \frac{-\lambda q^2 c^2}{\hbar^2} & 0 \\
	0 & 0 & 0 & 1 \\
	\frac{-\lambda q^2 c^2}{\hbar^2}  & 0 & \frac{-m^2 c^4 - q^2 c^2}{\hbar^2} & 0 \\
	\end{bmatrix} \hat{\phi} \equiv G \hat{\phi},
	\label{e:dynamical-matrix}
        \eea
        where  $\hat{\phi}=(\tilde{\phi}_{1},\tilde{\Pi}_{1},\tilde{\phi}_{2},\tilde{\Pi}_{2})^{\rm T}$. The matrix $G$ governs the dynamics of these field variables. In general, the dynamical matrix $G$ is non-Hermitian, i.e. $G \ne G^{\dagger}$. In fact, we find, $G^{\dagger} = \tau_2 G \tau_2$, where $\tau_2=\boldsymbol{1}_2 \otimes \sigma_2$; thus, $G$ is a pseudo-Hermitian matrix. Here, $\boldsymbol{1}_2$ is a $2\times2$ identity matrix and $\sigma_i$'s (for $i=1,2,3$) are Pauli matrices.

The eigenvalues of $G$ are $\pm (c/\hbar)\sqrt{c^2 m^2+q^2(1\pm \lambda)}$. For fixed values of $q$ and $m$, the eigenvalues of $G$ switch from real to complex conjugate pairs as we change $\lambda$ beyond $1+(cm/q)^2$.  When the eigenvalues become zero for some parameters, the eigenvectors also coincide. These values of parameters correspond to EPs in the parameter-space. The generalized $\mathcal{PT}$ symmetry of $G$ explains these features of the eigenvalues. In fact,  we find, $G^{*} = \tau_3 G \tau_3$ for $\tau_3=\boldsymbol{1}_2 \otimes \sigma_3$, which is in general, true for any $\mathcal{ PT}$ symmetric matrix. Let us consider,
	\bea
	[\mathcal{PT}, G] =0 \implies {\mathcal P} G^{*} = G {\mathcal P}.
	\eea 
Thus, $G$ is $\mathcal{PT}$ symmetric matrix when we identify the parity operator as $\mathcal{P} = \tau_3$. In general, a non-Hermitian (Hamiltonian) matrix admits a real spectrum for some values of the parameter even if the symmetry operator is not $\mathcal{PT}$, but instead is an anti-unitary operator $A$ satisfying the condition $A^{2r} = 1$ with $r$ odd \cite{Bender_2002_generalized}. In fact, this is known as the generalized $\mathcal{PT}$-symmetry. The unbroken regime of symmetry admits real spectrum while the spontaneous breaking of the symmetry leads to complex eigenvalues. To understand the symmetry of $G$ better, we notice that a combination of $\tau_3$ and complex conjugation operator $\kappa$ acts as the anti-unitary operator $A = \tau_3 {\kappa}$. Here, $\kappa$ acts as a time-reversal operator (${\cal T}$) for spin-less systems. The spectrum of the Hermitian Hamiltonian in Eq.~\ref{e:Coupled-FT-Hamiltonian} also drastically changes its features across the EPs.  
 
\section{Coupled scalar field theory on lattice}
\label{s:Lattice}

We discretize this field theory by placing it on a lattice  enabling the calculation of non-equilibrium heat transport using the lattice Green's functions. We use the redefinition, $y_{\alpha,n} = (cd^2/\hbar)^{1/2}\phi_{\alpha}(x=nd), p_{\alpha,n}  =(\hbar/c)^{1/2} \Pi_{\alpha}(x=nd)$ for $\alpha=1,2$ and $n=1,2,\dots N$. Here, $d$ is the lattice spacing. Thus, the effective low-energy (since $d$ introduces a UV cutoff) description of this field theory in  Eq.~\ref{e:Coupled-FT-Hamiltonian} is given by the Hamiltonian:
	\bea
	H &=&\sum_{\alpha =1}^2 \Big[\sum_{n =1}^{N} \Big(\frac{p_{\alpha,n}^2}{2m_0} + \frac{k_p}{2} y_{\alpha,n}^2 \Big) \cr
	&+&\sum_{n = 0}^{N} \frac{k}{2} (y_{\alpha,n+1} - y_{\alpha,n})^2 \Big] \cr
	&+& \lambda k \sum_{n = 0}^{N}(y_{1, n+1} - y_{1,n})(y_{2,n+1} - y_{2,n}).  
	\label{e:Network}
	\eea 
Here, $m_0 = \hbar/d c$, $k_p = (\hbar/d c)(m^2 c^4/\hbar^2)$ and $k = (\hbar/d c) (c/d)^2$. Physically, this Hamiltonian represents a network made by a pair of interconnected chains of harmonic oscillators (springs) representing two types of scalar fields. We have $[y_{\alpha,n},p_{\beta,n'}]= i \hbar \delta_{n n'} \delta_{\alpha \beta}$. The same kind of oscillators are connected to their nearest neighbors by spring constant $k$. The chains are coupled between each other through a coupling $\lambda k$. Moreover, each oscillator is locally pinned by a harmonic potential of strength $k_p$. We impose fixed boundary conditions, $y_{\alpha,0} = y_{\alpha,N+1} = 0$ for the network. The Hamiltonian in Eq.~\ref{e:Network} can be diagonalized by a series of transformations to normal modes: $y_{\alpha,n} = \sqrt{2/(N+1)} \sum_{q'=1}^{N} \tilde{y}_{\alpha,q'} \sin \left[\pi n q'/(N+1)\right],\quad \tilde{y}_{s/a,q'} = (\tilde{y}_{1, q'} \pm \tilde{y}_{2, q'})/\sqrt{2}$, and similar relations for $p_{\alpha,n}$. Thus, we get 
	\beq
	H = \sum_{q' = 1}^{N} \left[ \frac{\tilde{p}^2_{s,q'}}{2m_0}+ \frac{m_0 \Omega^2_{s,q'}}{2} \tilde{y}^2_{s,q'} + \frac{\tilde{p}^2_{a,q'}}{2m_0} + \frac{m_0 \Omega^2_{a,q'}}{2} \tilde{y}^2_{a,q'} \right],
	\eeq
where the frequencies $\Omega_{s,q'},\Omega_{a,q'}$ of symmetric and anti-symmetric normal modes are, $\Omega^2_{s/a,q'} = (k_p/m_0) + (4  k ( 1 \pm \lambda)/m_0) \sin^2\left[\pi q'/(2(N+1))\right]$. For $\lambda>\lambda_c\equiv 1 +(k_p/4k)\csc^2\left[\pi N/(2(N+1)) \right]$, $\Omega_{a,q'}$ can become imaginary implying the presence of inverted oscillator modes along with regular oscillators. Thus, the network can admit imaginary frequencies beyond a critical $\lambda$, which leads to an instability in the field theory. The evolution of some observables, e.g., the out-of-time order correlators (OTOC), is unbounded in time in the dynamically unstable regime for parameters with at least one inverted normal mode \cite{Ali_2020, Bhattacharyya_2021,Qu_2022}. In the absence of inverted normal modes, the time evolution of all observables including the OTOC is bounded in the dynamically stable regime.  

\section{Heat conduction}
\label{s:Heat}

We connect two ends of the oscillator network to heat baths kept at different temperatures $T_L,T_R$. We consider two different types of coupling between the network and the baths at the boundaries. We first take both the oscillators (fields) at any boundary being connected to a single bath with the same coupling strength. This leads to the equations of motion (EOM) for the oscillators at the boundaries in the following form for any general model of baths: 
	\bea
	m_0 \ddot{y}_{\alpha,1} &=& -k_p y_{\alpha,1} - k \Big[(2 y_{\alpha,1} - y_{\alpha,2}) \cr
	&+& \lambda (2 y_{\beta,1}  - y_{\beta,2}) \Big] \cr
	&+& \int_{-\infty}^{t} dt' \Sigma_{L}^{+}(t-t') \sqrt{2}y_{s,1}(t') + \eta_L,
	\label{e:eom1}\\  
	m_0 \ddot{y}_{\alpha,N} &=& -k_p y_{\alpha,N} - k \Big[ (2 y_{\alpha,N} - y_{\alpha,N-1}) \cr
	&+& \lambda ( 2 y_{\beta,N} -  y_{\beta,N-1}) \Big] \cr
	&+& \int_{-\infty}^{t} dt' \Sigma_{R}^{+}(t-t') \sqrt{2}y_{s, N}(t')  + \eta_{R}, 
	\label{e:eom2}
	\eea
where $\alpha \ne \beta, \alpha,\beta=1,2$, and the symmetric modes $y_{s, n}(t)=(y_{1, n}(t)+y_{2, n}(t))/\sqrt{2}$ for $n=1,2,\dots N$. Here, we assume the heat baths are connected to the network at infinite past. The noise terms $\eta_L(t),\eta_R(t)$ from the left and right heat baths are related to the self energies $\Sigma_{L}^{+},\Sigma_{R}^{+}$ of the corresponding baths through the fluctuation-dissipation relations as given in Eq.~\ref{e:FDR}. 

The heat current in the network can be evaluated by calculating the rate at which the heat bath at any boundary does work on the network \cite{Dhar_2006, Roy_2008}. Employing the continuity equation, we find the heat current $J_{\rm I}(t)$ at the left boundary as
	\beq
	J_{\rm I}(t)=-\dot{y}_{s,1}(t)\big(\sqrt{2}\eta_L(t)+\int_{-\infty}^tdt'2\Sigma_L^+(t-t') y_{s,1}(t')\big),
	\label{e:heat-current-J1}
	\eeq
which solely depends on the symmetric modes of the network. Such form of heat current appears due to the symmetric coupling of the heat baths with the network, which is also evident from the EOM for the oscillators at the boundaries in Eqs.~\ref{e:eom1} and \ref{e:eom2}. Since the frequency of the symmetric modes remains real-valued for all $\lambda$, these modes are bounded for all time. Therefore, $J_{\rm I}(t)$ is insensitive to the stability-to-instability transition in the network for symmetric connections to one heat bath at each boundary. We can find a steady state of $J_{\rm I}(t)$ both in the stable and unstable regime of the network. 

Next, we consider that each oscillator (field) at any boundary is connected to an individual heat bath with the same coupling strength. Both baths at any boundary are kept at the same temperature. Such coupling leads to the EOM for the oscillators at the boundaries in the following form for a general model of heat baths:
	\bea
	m_0 \ddot{y}_{\alpha,1} &=& -k_p y_{\alpha,1} - k \Big[ (2 y_{\alpha,1} - y_{\alpha,2}) \cr
	&+& \lambda (2 y_{\beta,1}  - y_{\beta,2}) \Big] \cr 		
	&+& \int_{-\infty}^{t} dt' \Sigma_{\alpha,L}^{+}(t-t') y_{\alpha,1}(t') + \eta_{\alpha,L},
	\label{e:eom3}\\  
	m_0 \ddot{y}_{\alpha,N} &=& -k_p y_{\alpha,N} - k \Big[ (2 y_{\alpha,N} - y_{\alpha,N-1}) \cr
	&+& \lambda ( 2 y_{\beta,N} -  y_{\beta,N-1}) \Big] \cr
	&+& \int_{-\infty}^{t} dt' \Sigma_{\alpha,R}^{+}(t-t') y_{\alpha, N}(t')  + \eta_{\alpha,R}, 
	\label{e:eom4}
	\eea
	where again $\alpha \ne \beta, \alpha,\beta=1,2$, and $\eta_{\alpha,L}~(\eta_{\alpha,R})$ and $\Sigma^{+}_{\alpha,L}~(\Sigma^{+}_{\alpha,R})$ are noise and self-energy of the two heat baths at the left (right) boundary. The heat current $J_{\rm II}(t)$ for this case from the continuity equation takes the following form at the left boundary:
	\bea
	J_{\rm II}(t)&=&-\sum_{\alpha=1,2}\dot{y}_{\alpha,1}(t)\big(\eta_{\alpha,L}(t)\nn\\&+&\int_{-\infty}^tdt'\Sigma_{\alpha,L}^+(t-t') 	y_{\alpha,1}(t')\big), 
	\label{e:heat-current-J2}
	\eea
which shows that both boundary oscillators appear explicitly in the heat current. Thus, $J_{\rm II}(t)$ depends on both the symmetric and anti-symmetric modes. Consequently, $J_{\rm II}(t)$ would not reach a steady state for the network in the unstable regime, as the anti-symmetric modes become unbounded over time. Nevertheless, $J_{\rm II}(t)$ acquires a steady-state value when the network is in the stable regime, since both the symmetric and anti-symmetric modes are bounded over time.

The qualitative difference in achieving a steady-state heat conduction due to different boundary conditions is one highlight of our present study. Similar differences in nature of heat conduction for different type of the boundary coupling have been shown earlier in heat transport through disordered harmonic lattices which attracted much attention \cite{Dhar2001,Roy2008, Chaudhuri2010}. We consider two different types of heat baths, namely, (a) baths modeled by semi-inﬁnite ordered harmonic chains (Rubin model of baths) \cite{rubin1971abnormal}, and (b) white noise baths \cite{casher1971heat}. The white noise in a bath is uncorrelated at any two different times signaling Markovian dynamics. The color noise in the Rubin model of bath are correlated for different times indicating non-Markovian feature.

The EOM for the oscillators' displacements in the bulk for $\alpha \ne \beta, \alpha,\beta=1,2$ and $n=2,3,\dots N-1$ are:
	\bea
	m_0 \ddot{y}_{\alpha,n} &=& -k_p y_{\alpha,n} - k \Big[(2 y_{\alpha,n} - y_{\alpha, n-1} - y_{\alpha,n+1}) \cr
	&+&   \lambda (2 y_{\beta,n} - y_{\beta,n+1} - y_{\beta,n-1}) \Big].
	\label{e:eom5}
	\eea
We solve these EOM in Eqs.~\ref{e:eom1}, \ref{e:eom2}, \ref{e:eom3}, \ref{e:eom4}, and \ref{e:eom5} using Fourier transforms of $y_{\alpha,n}(t), \eta_{b}(t),\eta_{\alpha, b}(t),\Sigma_{b}^{+}(t),\Sigma_{\alpha,b}^{+}(t)$ to frequency domain with $b=L,R$, e.g., 
	\beq
	\tilde{y}_{\alpha,n}(\omega) = \frac{1}{2\pi}\int_{-\infty}^{\infty}dt\,y_{\alpha,n}(t)e^{i\omega t}. 
	\label{e:FT1}
	\eeq
The self-energy $\tilde{\Sigma}_{b}^{+}(\omega) = i\gamma_{b}^o\omega$ for the white noise baths, and
	\beq
	\tilde{\Sigma}_{b}^{+}(\omega) = \frac{\gamma_{b}^2}{k_0} \Big[\Big(1-\frac{m_0 \omega^2}{2k_0} \Big) + i \Big(\frac{\omega \sqrt{m_0}}{\sqrt{k_0}}\sqrt{1-\frac{m_0 \omega^2}{4k_0}} \Big) \Big],
	\label{e:self-energy-Rubin-Bath}
	\eeq
for the Rubin baths, where $\gamma_L^o,\gamma_R^o,\gamma_L$, and $\gamma_R$ control the coupling of the respective boundary oscillators with the left and right Ohmic and Rubin baths, respectively. This leads to $\tilde{\Sigma}_{\alpha,b}^{+}(\omega)=\tilde{\Sigma}_{b}^{+}(\omega)$ for $\alpha=1,2$. Here, $2\sqrt{k_0}$ is the finite band-width of the Rubin baths in comparison to infinite band width for the white noise baths. We further introduce the symmetric and anti-symmetric modes as $\tilde{y}_{s/a,n}(\omega)=(\tilde{y}_{1,n}(\omega)\pm\tilde{y}_{2,n}(\omega))/\sqrt{2}$. 

\section{Single bath at each boundary}
\label{s:Single-bath}

For a single bath at each boundary of the network, the EOM in Eqs.~\ref{e:eom1} and \ref{e:eom2} get decoupled in terms of these symmetric and anti-symmetric modes as,
	\bea
	z_{L+}\tilde{y}_{s,1}(\omega)-k(1+\lambda)\tilde{y}_{s,2}(\omega)  &=& \sqrt{2}\tilde{\eta}_{L}(\omega),
	\label{e:fm1}\\
	z_{-}\tilde{y}_{a,1}(\omega)-k(1-\lambda)\tilde{y}_{a,2}(\omega)  &=& 0,
	\label{e:fm2}\\
	z_{R+}\tilde{y}_{s,N}(\omega)-k(1+\lambda)\tilde{y}_{s,N-1}(\omega)  &=& \sqrt{2}\tilde{\eta}_{R}(\omega),
	\label{e:fm3}\\
	z_{-}\tilde{y}_{a,N}(\omega)- k(1-\lambda)\tilde{y}_{a,N-1}(\omega)  &=& 0,
	\label{e:fm4}
	\eea
where $z_{b+}=-m_0 \omega^2 + k_p + 2 k(1+\lambda) - 2 \tilde{\Sigma}^+_{b}(\omega)$, and $z_{-}=-m_0 \omega^2 + k_p + 2 k(1-\lambda)$ for $b=L,R$. Here,  $\tilde{\eta}_{b}(\omega)=(1/2\pi)\int_{-\infty}^{\infty}dt\,\eta_{b}(t)e^{i\omega t}$. It is also possible to apply the symmetric and anti-symmetric mode transformation first in the time-domain, and then perform the Fourier transforms to these modes. The application of Fourier transforms is only allowed for these modes when they are in steady state. Thus, the aforementioned Eqs.~\ref{e:fm1} and \ref{e:fm3} for the symmetric modes are valid in both the stable and unstable regime as these modes with real frequencies remain bounded at long time even in the unstable regime of the network.  However, the Eqs.~\ref{e:fm2} and \ref{e:fm4} are only correct for the network in the stable regime. We can formally write the solution of the symmetric modes as $\tilde{Y}_s(\omega) = G^{+}_s(\omega)\tilde{\eta}_s(\omega)$, where $\tilde{Y}_s$ and $\tilde{\eta}_s$ are column vectors of dimension $N$: $\tilde{Y}_s = (\tilde{y}_{s,1},\tilde{y}_{s,2}, \ldots \tilde{y}_{s,N})^{T} \quad \text{and} \quad \tilde{\eta}_s = (\sqrt{2}\tilde{\eta}_L, 0, \ldots, 0, \sqrt{2}\tilde{\eta}_R)^T$, and the retarded Green's function $G^{+}_s= Z_{s}^{-1}/\lambda_{+}$ with $\lambda_+= k(1+\lambda)$. Here, $Z_{s}$ is a $N \times N$ symmetric tridiagonal matrix whose offdiagonal elements are $-1$ and diagonal elements excluding the first and last ones are $(-m_0 \omega^2 + k_p)/\lambda_+ + 2$. The first and last diagonal elements of $Z_{s}$ are $z_{L+}/\lambda_+$ and $z_{R+}/\lambda_+$, respectively. 

The steady-state heat current $\bra J_{\rm I}\ket$ after noise averaging over $J_{\rm I}(t)$ in Eq.~\ref{e:heat-current-J1} reads 
	\beq
	\bra J_{\rm I} \ket = \int_{-\infty}^{\infty} \frac{d\omega \, 4\hbar \omega}{\pi} |[G^+_s(\omega)]_{1,N}|^2 \tilde{\Gamma}_L(\omega) \tilde{\Gamma}_R(\omega) (f_L -f_R),
	\label{e:heatcur-RG-J1}
	\eeq
where $\tilde{\Gamma}_{b}(\omega)$ are the imaginary part of $\tilde{\Sigma}^+_{b}(\omega)$, and $f_{b} \equiv f(\omega, T_{b})=1/(e^{\hbar \omega/K_BT_{b}}-1)$ is the Bose distribution function for phonons of the left and right ($b=L,R$) heat baths at temperature $T_b$. In deriving the current formula, we employ the fluctuation-dissipation relations in the frequency domain:
	\beq
	\bra \tilde{\eta}_{L}(\omega) \tilde{\eta}_{L}(\omega') \ket = \frac{{\rm Im}[\tilde{\Sigma}^+_L(\omega)]\hbar}{\pi}(1+f_L) \delta(\omega + \omega'), 
	\label{e:FDR}
	\eeq
and a similar relation for the right heat bath. These fluctuation-dissipation relations are true for both these white noise and Rubin baths with an appropriate self-energy contribution. We can derive an explicit expression for the heat current from Eq.~\ref{e:heatcur-RG-J1} in the linear response regime with an applied temperature difference $\Delta T=T_L-T_R \ll T \equiv (T_L+T_R)/2$. Expanding the phonon distributions $f(\omega, T_{b})$ about the mean temperature $T$, we get
	\bea
	\bra J_{\rm I} \ket &=&\frac{{4}k_B\Delta T}{\pi}\int_{-2\sqrt{k_0}}^{2\sqrt{k_0}} d\omega \Big[\left( \frac{\hbar \omega}{2 k_B T} \right)^2  {\rm csch}^2\left(\frac{\hbar \omega}{2 k_B T} \right) \nn\\ &&\tilde{\Gamma}_L(\omega) \tilde{\Gamma}_R(\omega)|[G^+_s(\omega)]_{1,N}|^2 \Big].
	\label{e:heatcur-RG-J1-lin-res}
	\eea
We can find an analytical expression for heat current in the high-temperature classical regime by taking $(\hbar \omega/2 k_B T)^2{\rm csch}^2(\hbar \omega/2 k_B T) \to 1$. We get a relatively compact formula of the steady-state classical heat current $\bra J_{\rm I}^{cl} \ket$ by taking $m_0=1,k_p=0, k(1+\lambda)=k_0$, and $\gamma_L=\gamma_R=\gamma$ for the Rubin baths:
	\beq
	\bra J_{\rm I}^{cl} \ket = \frac{4k_B\Delta T}{\pi}\int_{-2\sqrt{k_0}}^{2\sqrt{k_0}} d\omega \, |[G^+_s(\omega)]_{1,N}|^2 \frac{\omega^2\gamma^4}{k_0^3}\Big(1-\frac{\omega^2}{4k_0}\Big).
	\label{e:heatcur-RG-J1-clas}
	\eeq
The required Green's function element can be found as $|[G^+_s(\omega)]_{1,N}|=1/(\lambda_+|{\rm det} Z_s|)$. Applying the methods from Refs.~\cite{Hu_1996,Roy_2008} for tridiagonal symmetric matrix as explained in detail in Appendix~\ref{a:Single-bath-at-each-boundary-Rubin-Model}, we can find a simple form of $|{\rm det} Z_s|$ in the large $N$ limit with a parametrization $\omega^2=4k_0\sin^2(q/2)$ to write
	\bea
	\bra J_{\rm I}^{cl} \ket &=& \frac{k_B\Delta T \Omega \sqrt{k_0}}{\pi}\int_{0}^{\pi} dq \frac{\cos(q/2)\sin^2q}{\Upsilon-		\Omega\cos(2q)}\nn\\
	&=& \frac{k_B\Delta T \sqrt{k_0}}{\pi} \Big(1  \cr
	&-& \sum_{s=\pm1}\frac{s(\Upsilon-\Omega)\tan^{-1}\big[\frac{2\sqrt{\Omega}}				{\sqrt{-2\Omega+s\sqrt{2\Omega\tilde{\Upsilon}}}}\big]}{2\sqrt{2\tilde{\Upsilon}}											\sqrt{-2\Omega+s\sqrt{2\Omega\tilde{\Upsilon}}}} \Big),
	\label{e:heatcur-RG-J1-clas-analy-exp}
	\eea
where $\Upsilon = k_0^4-2k_0^2\gamma^2+4\gamma^4, \Omega = 2k_0^2\gamma^2$ and $\tilde{\Upsilon}=\Upsilon+\Omega$. The system-size independent thermal current is a signature of ballistic heat transport in the harmonic networks of oscillators. The expression in Eq.~\ref{e:heatcur-RG-J1-clas-analy-exp} is valid for different values of $\lambda$ giving rise to stable and unstable phases.

\begin{figure}[h!]
	\includegraphics[width=\linewidth, height=0.7\linewidth]{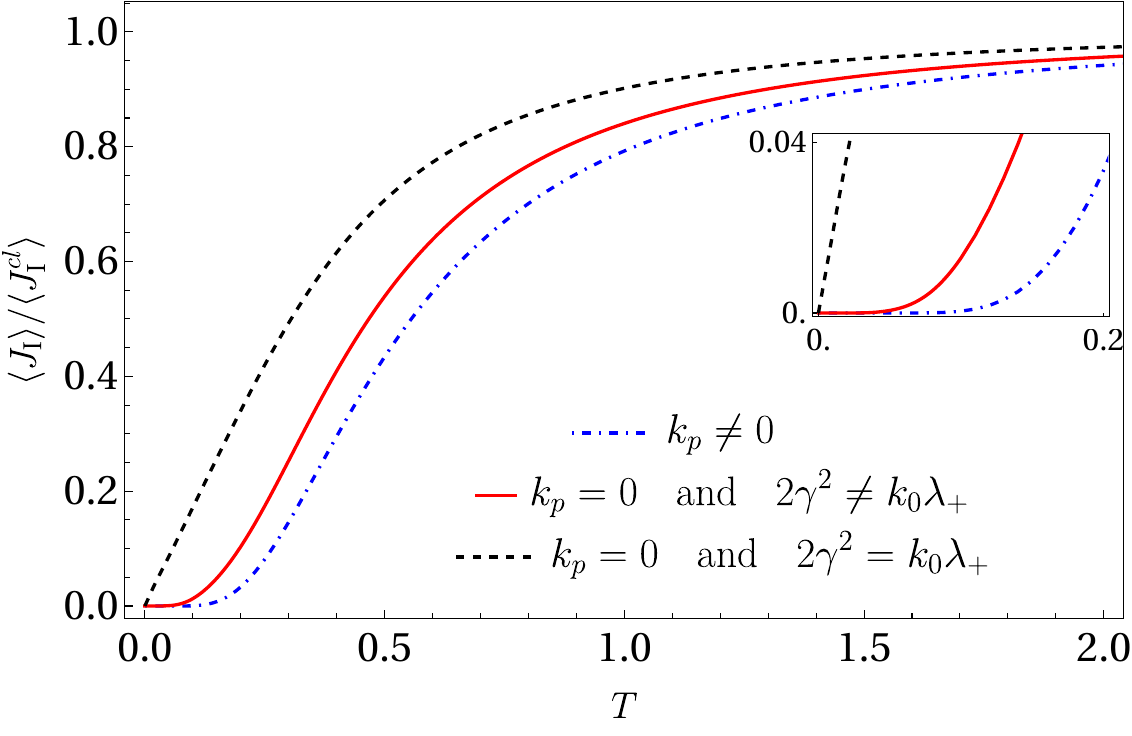}
  	\caption{Temperature dependence of the scaled heat current $\bra J_{\rm I} \ket/\bra J_{\rm I}^{cl} \ket$ with a 			single bath at each boundary for different pinning $k_p$ and bath coupling $\gamma$. The inset shows features at low temperature for three different parameter regimes (see text). The temperature is in units of $(\hbar\sqrt{k/m_0})/k_B$.}
	\label{f:Temp-dependence-J1}
	\end{figure}

We also derive an analytical formula of the thermal current in the linear response regime at high temperature for arbitrary $m_0, k_p, k$, and $\lambda$. We give its derivation in Appendix~\ref{a:Single-bath-at-each-boundary-Rubin-Model}. Next, we discuss the features of $\bra J_{\rm I} \ket$ at low-temperature quantum regime. The $T$-dependence of $\bra J_{\rm I} \ket$ can be extracted from the following expression by considering that only low-frequency (or wavevector) modes contribute in heat current when $T\to 0$. The quantum heat current
	\bea
	\bra J_{\rm I} \ket &=&\frac{k_B\Delta T}{\pi}\int_{-\pi}^{\pi} dq |\frac{d\omega}{dq}||[G^+_s(q)]_{1,N}|^2 					\frac{4 m_0 \omega^2\gamma^4}{k_0^3}\nn\\ &\times&\Big(1-\frac{ m_0\omega^2}{4k_0}\Big)\left( \frac{\hbar \omega}{2 k_B T} \right)^2  {\rm csch}^2\left(\frac{\hbar \omega}{2 k_B T} \right),
	\label{e:heatcur-RG-J1-lin-res}
	\eea
where $\omega^2=(k_p+4k(1+\lambda)\sin^2(q/2))/m_0$. We determine $T$-dependence of $\bra J_{\rm I} \ket$ by finding the leading $q$-dependence of $|[G^+_s(q)]_{1,N}|^2$ in the large $N$ limit, and performing a scaling analysis of Eq.~\ref{e:heatcur-RG-J1-lin-res}. In the pinned (massive scalar fields) case when $k_p\ne 0$, for Rubin baths, we find $\bra J_{\rm I} \ket \sim (e^{-\hbar \omega_0/(k_BT)})/T^{1/2}$  with $\omega_0 = \sqrt{k_p/m_0}$ (see Appendix~\ref{a:Single-bath-at-each-boundary-Rubin-Model}). For unpinned (massless scalar fields) case when $k_p=0$, $\bra J_{\rm I} \ket \sim T^3$ when $2 \gamma^2 \neq k_0 \lambda_+$ and $\bra J_{\rm I} \ket \sim T$ when $2 \gamma^2 = k_0 \lambda_+$ as shown in Appendix~\ref{a:Single-bath-at-each-boundary-Rubin-Model}. In Fig.~\ref{f:Temp-dependence-J1}, we plot the scaled heat current $\bra J_{\rm I} \ket/\bra J_{\rm I}^{cl} \ket$  with temperature when (i) $k_p/k= 0.5, \gamma/k= 0.2$, (ii) $k_p = 0, \gamma/k= 0.2$ for $2 \gamma^2 \neq k_0 \lambda_+$, and (iii) $k_p = 0, \gamma/ k = \sqrt{2.4}$ satisfying $2 \gamma^2=k_0 \lambda_+$, with $k_0/k = 4, \lambda = 0.2$ and $m_0= \hbar = k_B = 1$. The inset shows the low-temperature features of $\bra J_{\rm I} \ket/\bra J_{\rm I}^{cl} \ket$. In the numerical analysis, we set $\Delta T = 0.001$. The above predicted $T$-dependencies of $\bra J_{\rm I} \ket$ match with the numerical results shown in Fig.~\ref{f:Temp-dependence-J1}, for these three different cases. 

\section{Two baths at each boundary}
\label{s:Two-baths}

The EOM in Eqs.~\ref{e:eom3} and \ref{e:eom4} for the boundary oscillators connected to two different baths with the same coupling strength read in terms of the symmetric and anti-symmetric modes in the frequency domain as:
	\bea
	\tilde{z}_{L+}\tilde{y}_{s,1}(\omega)-k(1+\lambda)\tilde{y}_{s,2}(\omega)  &=& \tilde{\eta}_{s,L}(\omega),
	\label{e:fm5}\\
	\tilde{z}_{L-}\tilde{y}_{a,1}(\omega)-k(1-\lambda)\tilde{y}_{a,2}(\omega)  &=& \tilde{\eta}_{a,L}(\omega),
	\label{e:fm6}\\
	\tilde{z}_{R+}\tilde{y}_{s,N}(\omega)-k(1+\lambda)\tilde{y}_{s,N-1}(\omega)  &=& \tilde{\eta}_{s,R}(\omega),
	\label{e:fm7}\\
	\tilde{z}_{R-}\tilde{y}_{a,N}(\omega)- k(1-\lambda)\tilde{y}_{a,N-1}(\omega)  &=& \tilde{\eta}_{a,R}(\omega),
	\label{e:fm8}
	\eea
where $\tilde{z}_{b\pm}=-m_0 \omega^2 + k_p + 2 k(1\pm\lambda) - \tilde{\Sigma}^+_{b}(\omega)$, $\tilde{\eta}_{s,b}(\omega)=(\tilde{\eta}_{1,b}(\omega)+\tilde{\eta}_{2,b}(\omega))/\sqrt{2}$, and $\tilde{\eta}_{a,b}(\omega)=(\tilde{\eta}_{1,b}(\omega)- \tilde{\eta}_{2,b}(\omega))/\sqrt{2}$ for $b=L,R$.  Similar to the case of single bath at each boundary, the application of Fourier transforms to frequency domain is only allowed for these modes in the steady-state, which occurs for the network in the dynamically stable regime. The EOM for the bulk oscillators in Eq.~\ref{e:eom5} remain the same for this case. We apply Fourier transformation after switching to these symmetric and anti-symmetric modes. We can again formally write the solution of both the symmetric and anti-symmetric modes as $\bar{Y}_{s/a}(\omega) = \bar{G}^{+}_{s/a}(\omega)\bar{\eta}_{s/a}(\omega)$, where $\bar{Y}_{s/a}$ and $\bar{\eta}_{s/a}$ are column vectors of dimension $N$: $\bar{Y}_{s/a} = (\tilde{y}_{s/a,1},\tilde{y}_{s/a,2}, \ldots \tilde{y}_{s/a,N})^{T} \quad \text{and} \quad \bar{\eta}_{s/a} = (\tilde{\eta}_{s/a,L}, 0, \ldots, 0, \tilde{\eta}_{s/a,R})^T$, and the retarded Green's functions $\bar{G}^{+}_{s}= \bar{Z}_{s}^{-1}/\lambda_{+}$ and $\bar{G}^{+}_{a}= \bar{Z}_{a}^{-1}/\lambda_{-}$ with $\lambda_{\pm}= k(1\pm\lambda)$. Here, $\bar{Z}_{s,a}$ are $N \times N$ symmetric tridiagonal matrices whose offdiagonal elements are $-1$ and diagonal elements excluding the first and last ones are $(-m_0 \omega^2 + k_p)/\lambda_{+,-} + 2$. The first and last diagonal elements of $\bar{Z}_{s}$ are  $\tilde{z}_{L+}/\lambda_+$ and $\tilde{z}_{R+}/\lambda_+$, respectively, and those for $\bar{Z}_{a}$ are $\tilde{z}_{L-}/\lambda_-$ and $\tilde{z}_{R-}/\lambda_-$,  respectively. 

\begin{figure}[h!]
  	\includegraphics[width=\linewidth, height=0.7\linewidth]{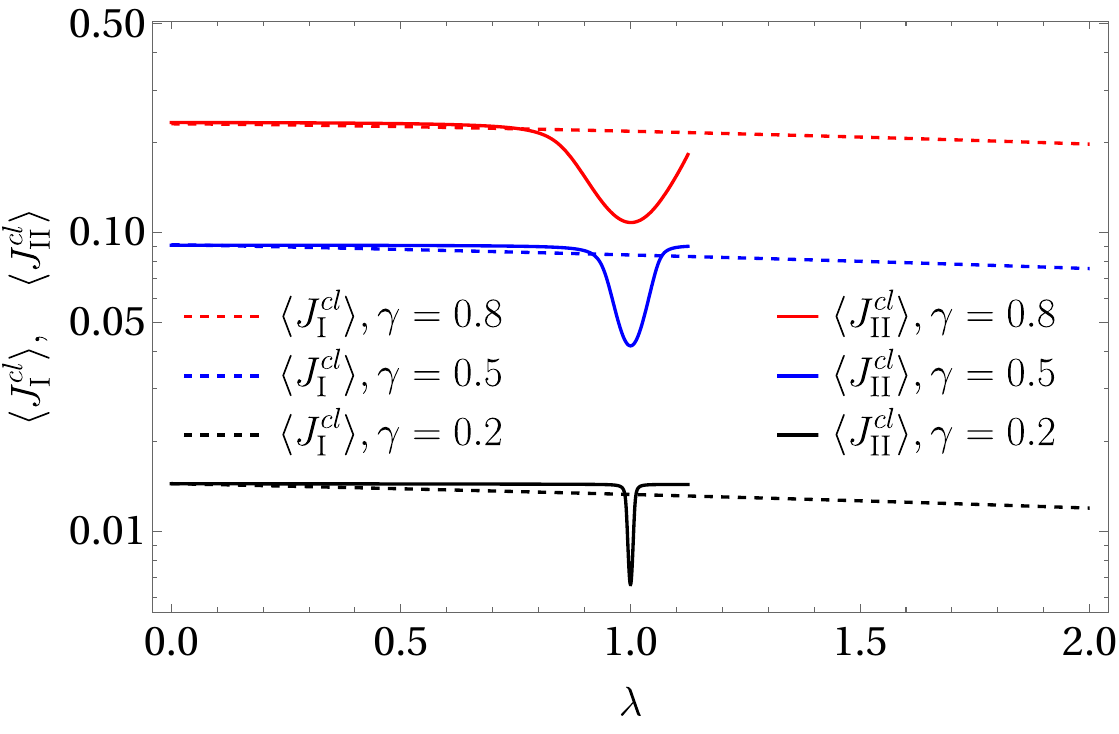}
  	\caption{Comparison between the high-temperature classical heat currents $\bra J_{\rm I}^{cl}\ket$ (dashed lines) and $\bra J_{\rm II}^{cl}\ket$ (solid lines), (the $y$-axis represents the heat currents multiplied by the common factor $\pi/ (k_B \Delta T)$ in the $\log$ scale), with an increasing coupling $\lambda$ between the fields for different bath coupling strengths $\gamma~(=0.8,0.5,0.2$ for upper to lower plots). The parameter $\gamma$ is in units of $k$. We set $k_p = 0.5 k, k_0 = 4 k$ and $m_0 =  k = \hbar = k_B = 1$ in the numerical analysis. Here, $\bra J_{\rm II}^{cl}\ket$ can be defined properly only upto the critical coupling $\lambda_b \approx 1 + (k_p/4 k)$.}
	\label{f:compare-classical-J1-J2}
	\end{figure}

\begin{figure}[h!]
  	\includegraphics[width=\linewidth, height=0.7\linewidth]{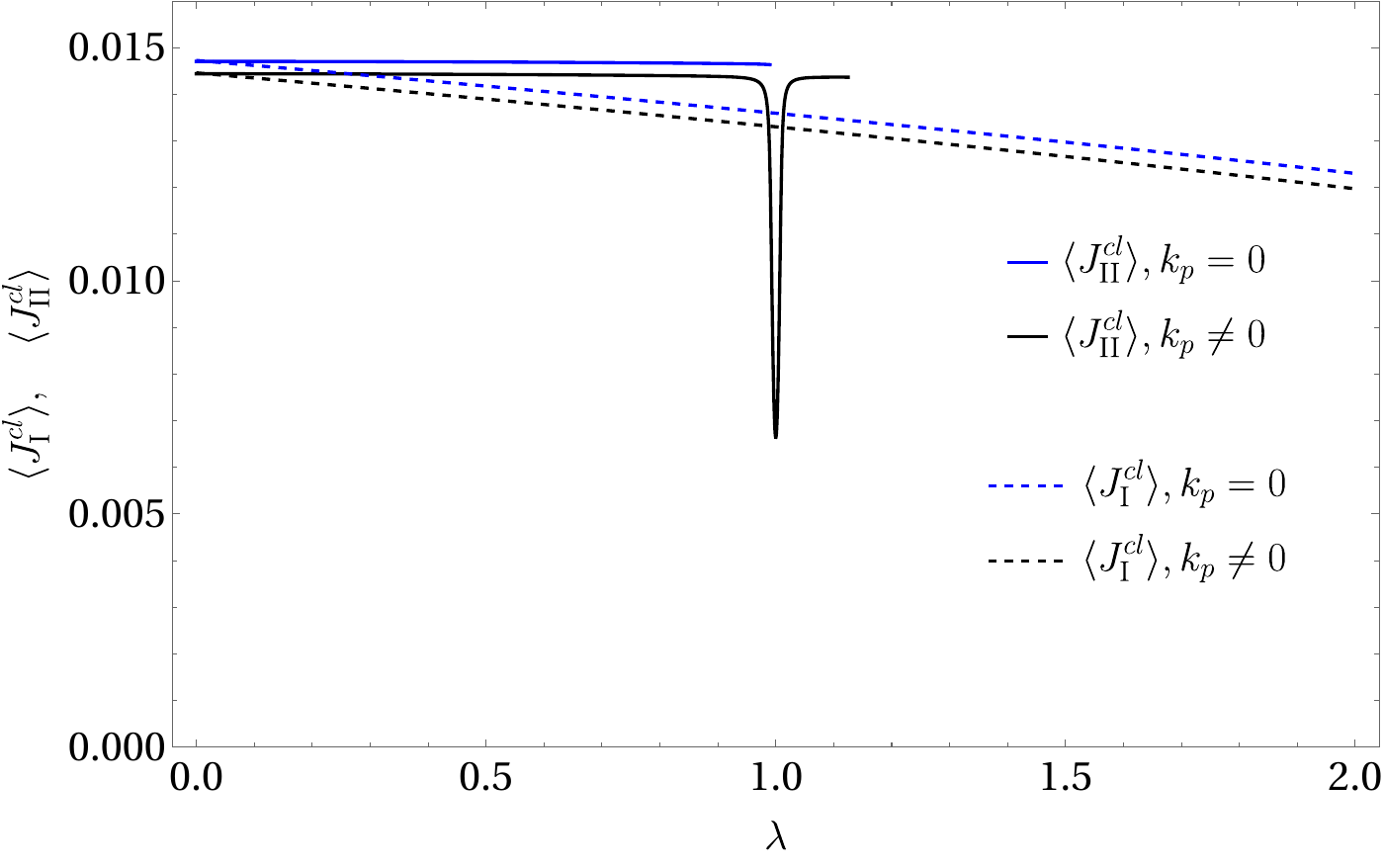}
  	\caption{Comparison between the high-temperature classical heat currents $\bra J_{\rm I}^{cl}\ket$ (dashed lines) and $\bra J_{\rm II}^{cl}\ket$ (solid lines), (the $y$-axis represents the heat currents multiplied by the common factor $\pi/ (k_B \Delta T)$), with an increasing coupling $\lambda$ between the fields for fixed $\gamma = 0.2 k$ and for both zero and nonzero $k_p$ values. We set either $k_p =0$ (blue) or $k_p = 0.5 k$ (black) and   $k_0 = 4 k, m_0 =  k = \hbar = k_B = 1$ in the numerical analysis. Here, $\bra J_{\rm II}^{cl}\ket$ is again defined properly upto a critical coupling strength $\lambda_a = 1 - (\gamma^2/k k_0)$ and $\lambda_b \approx 1 + (k_p/4 k)$ for $k_p =0$ and $k_p \neq 0$, respectively. Notice that the value of current is slightly large for $k_p =0$ in both cases.}
	\label{f:compare-classical-J1-J2-for-kp-zero-and-nonzero}
	\end{figure}

By performing the noise averaging of $J_{\rm II}$ in Eq.~\ref{e:heat-current-J2}, we find the steady-state heat current $\bra J_{\rm II} \ket$ in the stable regime of the network as:  
	\bea
	\bra J_{\rm II} \ket &=&\int_{-\infty}^{\infty}d\omega \Big[\frac{\hbar \omega}{\pi}(|[\bar{G}^+_s(\omega)]_{1,N}|^2+|[\bar{G}^+_a(\omega)]_{1,N}|^2) \cr
	&\times& \tilde{\Gamma}_L(\omega)\tilde{\Gamma}_R(\omega) (f_L -f_R) \Big],
	\label{e:heatcur-RG-J2}
	\eea
where we have applied the following fluctuation-dissipation relations for the noise averaging:
	\beq
	\bra \tilde{\eta}_{\alpha,L}(\omega) \tilde{\eta}_{\beta,L}(\omega') \ket = \delta_{\alpha,\beta} \frac{{\rm Im}[\tilde{\Sigma}^+_L(\omega)]\hbar}{\pi}(1+f_L) \delta(\omega + \omega'). 
	\label{e:FDR1}
	\eeq
We again work in the linear response regime for $\Delta T \ll T$ to derive an analytical expression of $\bra J_{\rm II} \ket$. Expanding the phonon distributions $f(\omega, T_b)$ about the mean temperature $T$, from Eq.~\ref{e:heatcur-RG-J2} we obtain
	\bea
	\bra J_{\rm II} \ket &=&\frac{k_B\Delta T}{\pi}\Big(\int_{-\omega_{1+}}^{\omega_{1+}} d\omega |[\bar{G}^+_s(\omega)]_{1,N}|^2\nn\\&+&\int_{-\omega_{1-}}^{\omega_{1-}} d\omega |[\bar{G}^+_a(\omega)]_{1,N}|^2\Big) \Big[\tilde{\Gamma}_L(\omega) \tilde{\Gamma}_R(\omega) \nn\\&\times&\left( \frac{\hbar \omega}{2 k_B T} \right)^2  {\rm csch}^2\left(\frac{\hbar \omega}{2 k_B T} \right) \Big],
	\label{e:heatcur-RG-J2-lin-res}
	\eea
where $\omega_{1\pm}=\sqrt{(k_p+4k(1\pm\lambda))/m_0}$. In the high-temperature classical regime for the Rubin baths, using  Eq.~\ref{e:self-energy-Rubin-Bath}, we derive from Eq.~\ref{e:heatcur-RG-J2-lin-res}
	\bea
	\bra J_{\rm II}^{cl} \ket &=&\frac{k_B\Delta T}{\pi}\Big(\int_{-\omega_{1+}}^{\omega_{1+}} d\omega |[\bar{G}				^+_s(\omega)]_{1,N}|^2\nn\\&+&\int_{-\omega_{1-}}^{\omega_{1-}} d\omega |[\bar{G}^+_a(\omega)]_{1,N}|^2\Big)				\frac{m_0 \omega^2\gamma^4}{k_0^3}\Big(1-\frac{m_0 \omega^2}{4k_0}\Big).  \nn \\ 
	\label{e:heatcur-RG-J2-clas}
	\eea
        We find the analytical expression of $[\bar{G}^+_{s/a}(\omega)]_{1,N}$ in the large $N$ limit following Refs.~\cite{Hu_1996,Roy_2008}, and then perform the integrals in Eq.~\ref{e:heatcur-RG-J2-clas} to write long analytical formulas for the symmetric and anti-symmetric mode contributions to $\bra J_{\rm II}^{cl} \ket$ in Eq.~\ref{e:analytical-formula-for-sym-antsym-currents} of Appendix~\ref{b:Two-baths-at-each-boundary-Rubin-Model}. In addition, $\bra J_{\rm II}^{cl} \ket$ is independent of $N$ indicating ballistic heat transport in the harmonic network for such a bath coupling. We compare $\bra J_{\rm II}^{cl} \ket$ to $\bra J_{\rm I}^{cl} \ket$ with increasing $\lambda$ in the stable regime of the network for similar parameters of the network and the bath coupling strengths $\gamma$ in Figs.~\ref{f:compare-classical-J1-J2},\ref{f:compare-classical-J1-J2-for-kp-zero-and-nonzero}. In general, $\bra J_{\rm II}^{cl} \ket$ is slightly larger than $\bra J_{\rm I}^{cl} \ket$ for smaller $\lambda$s till a specific value of $\lambda$ (denoted as $\lambda_{a}$) depending on the value of $\gamma$ (see Fig.~\ref{f:compare-classical-J1-J2-for-kp-zero-and-nonzero}). Two baths at each edge facilitate the thermalization of the network better than a single bath at each edge, thus resulting in higher heat transport in the two baths case. We further observe from Eq.~\ref{clcurr2} that the anti-symmetric mode contribution to $\bra J_{\rm II}^{cl} \ket$ vanishes at $\lambda = 1$. 

        For a given value of $\gamma$, we determine the specific coupling strength $\lambda_a$, at which $\bra J_{\rm II}^{cl} \ket$ exhibits a non-analytic behavior when $k_p =0$. The value of $\lambda_a$ is obtained analytically by solving the equation $\gamma^3 + k k_0 \gamma (\lambda -1) = 0$, which follows from the condition $\xi_{-}^a =1$ (see Eq.~\ref{e:xi-pm-sym-and-anti-sym} in  Appendix~\ref{b:Two-baths-at-each-boundary-Rubin-Model}), implying $\lambda_a = 1 - (\gamma^2/ k k_0)$. At $\lambda_a$, the analytical expression for the classical current (see Eq.~\ref{e:analytical-formula-for-sym-antsym-currents} of Appendix~\ref{b:Two-baths-at-each-boundary-Rubin-Model}) diverges and $\bra J_{\rm II}^{cl} \ket$ is properly defined in the stable regime only upto $\lambda_a$ as shown in  Fig.~\ref{f:compare-classical-J1-J2-for-kp-zero-and-nonzero}. In other words, the anti-symmetric part of $\bra J_{\rm II}^{cl} \ket$ is properly defined only in the regime $0 < \lambda < \lambda_a$. Thus, the bath coupling $\gamma$ reduces the critical coupling strength of the open network  by a factor $\gamma^2/k k_0$ compared to that of the closed network $\lambda_c = 1$ (for $k_p = 0$ and $N \to \infty$). For $k_p \neq 0$,  $\bra J_{\rm II}^{cl} \ket$ does not exhibit any such non-analytic behavior. Nevertheless, $\bra J_{\rm II}^{cl} \ket$ falls around $\lambda_a$ even when $k_p \neq 0$ due to decreasing contribution of the anti-symmetric modes to $\bra J_{\rm II}^{cl} \ket$  with increasing $\lambda$. 

We finally discuss the features of $\bra J_{\rm II} \ket$ at low temperature quantum regime. It can be determined from $\bra J_{\rm II} \ket$ in Eq.~\ref{e:heatcur-RG-J2-lin-res}. We could not calculate the integration in Eq.~\ref{e:heatcur-RG-J2-lin-res} for an arbitrary temperature. Nevertheless, we can extract $T$-dependence of $\bra J_{\rm II} \ket$ by finding leading $\omega$-dependence of $|[\bar{G}^+_{s/a}(\omega)]_{1,N}|^2$ in the large $N$ limit, and performing a scaling analysis of Eq.~\ref{e:heatcur-RG-J2-lin-res}. For Rubin baths, we again find in Appendix~\ref{b:Two-baths-at-each-boundary-Rubin-Model}, $\bra J_{\rm II} \ket \sim (e^{-\hbar \omega_0/(k_BT)})/T^{1/2}$ with $\omega_0 = \sqrt{k_p/m_0}$ for pinned (massive scalar fields) case when $k_p\ne 0$, which is similar to that for single bath at each boundary and for the white noise baths (see Appendix~\ref{c:Two-baths-at-each boundary-white-noise}). For unpinned (massless scalar fields) case when $k_p=0$, $\bra J_{\rm II} \ket \sim T$ if either $\gamma^2 = k_0 \lambda_+$ or $\gamma^2 = k_0 \lambda_-$, and $\bra J_{\rm II} \ket \sim T^3$ when $\gamma^2 \ne k_0 \lambda_+ \ne k_0 \lambda_-$  as shown in Appendix~\ref{b:Two-baths-at-each-boundary-Rubin-Model}. The corresponding white noise bath case is discussed in Appendix~\ref{c:Two-baths-at-each boundary-white-noise}.

\section{Summary and outlook}
\label{s:Summary}

We have explored heat conduction through a network of different types of oscillators representing a coupled scalar field theory driven by heat baths with different temperatures placed at the boundaries. The dynamical matrix of the network is non-Hermitian and undergoes a stability-to-instability transition at EPs as the coupling strength between the scalar fields (or different types of oscillators) increases. The unstable regime of the dynamical matrix is marked by the emergence of inverted oscillator modes in the network. Consequently, the open network never obtains a steady state in this unstable regime, and the heat current is unbounded for arbitrary bath coupling at the boundaries. In this work, we engineered a unique bath coupling where a single bath is connected to two fields at each edge with the same strength, leading to a finite steady-state heat conduction in the network for any arbitrary coupling strength between the fields. This occurs as the symmetric modes carry the heat, and the heat current is independent of the asymmetric modes, which becomes unbounded in the unstable regime. We also studied the coupling of two fields to two separate baths at each boundary. In this case, heat is carried by both the symmetric and anti-symmetric modes; thus, the heat current is well-behaved only in the stable regime.

We compared heat conduction through the network for collective versus individual coupling of fields to baths at the boundaries. Individual bath coupling of a boundary oscillator is primarily used in studying heat conduction through ordered or disordered networks \cite{Chaudhuri2010}. We have demonstrated an exciting distinction between different types of bath coupling, and these differences can be significant in the presence of stronger coupling between the network's modes and the baths. In contrast to existing numerical studies in complex networks with multiple modes \cite{Dhar2008, Chaudhuri2010}, our explicit analytical formulae in the thermodynamic limit can further aid in understanding the role of various parameters in controlling heat transport in networks with different bath coupling and bath models. One immediate extension of our study could be to explore how heat conduction and the stability-to-instability transition vary due to different types of disorder in the parameters of the network.

Further, we have assumed the fields or oscillator displacements are small for our quadratic modeling of the system Hamiltonian. However, such an assumption breaks down in the unstable regime when the displacement becomes unbounded, and we need to incorporate nonlinear terms in the Hamiltonian \cite{Liu2014}. We have observed that nonlinear interaction can regularize the heat current in the unstable regime, even for general system-bath couplings.

The stability-to-instability transition at the EPs of the non-Hermitian dynamical matrix in our present model is unrelated to a topological phase transition. Nevertheless, recent studies with various quadratic Hermitian bosonic models, whose dynamical matrices resemble non-Hermitian Hamiltonians, have revealed exciting topological phases and phase transitions \cite{McDonald_2018, Lieu2018, Wang2019, Flynn_2020, Flynn_2021}. Investigating heat conduction in such quadratic Hermitian bosonic models could be applied to identify topological phases and phase transitions in the related non-Hermitian Hamiltonians \cite{NB2022, Krishanu2021}. Identifying topological phases and phase transitions in non-Hermitian Hamiltonians is a key challenge of significant interest \cite{PhysRevX.8.021066,kawabata_symmetry_2019,wang_topological_2021,ritutopo2022,nehra2023anomalous}. 
Studying heat transport in this direction is significant and constitutes one of our future goals.

\vspace{0.5cm}
{\it Acknowledgements:} We would like to thank A. Clerk and K. Roychowdhury for the insightful discussions. T.R.V. extends gratitude to R. Bag, M. Eledath, K. Estake, G. Krishnaswami, V. Kumar, R. Nehra, and I. Santra for their valuable suggestions. 

\appendix

\section{Single Rubin bath at each boundary}
\label{a:Single-bath-at-each-boundary-Rubin-Model}

In this appendix, we consider that both types of oscillators at the left or the right end of the network are connected to a single Rubin bath with the same coupling strength. Thus, the couplings between the network and the baths are modeled by the Hamiltonian: 
	\beq
	H_{c1} = -\gamma_L y_{L,1}(y_{1,1} + y_{2,1}) - \gamma_R y_{R,1}(y_{1,N}+ y_{2,N}).
	\label{e:system-bath-coupling-single-bath-Rubin}	
	\eeq 
Here, $y_{L,n}$ and $y_{R,n}$ represent the left and right bath variables, and $\gamma_L$ and $\gamma_R$ are the coupling strengths between the left and right ends of the system and the corresponding baths, respectively. The Rubin baths are modeled as a chain of finite number ($N_b$) of harmonic oscillators with mass $m_0$ and spring constant $k_0$ as given by the Hamiltonian:
	\beq
	H_{b} = \sum_{n = 1}^{N_b} \frac{p_{b,n}^2}{2 m_0} + \sum_{n=0}^{N_b}\frac{k_{0}(y_{b,n} - y_{b,n+ 1})^2}{2}, 
	\label{e:bath-Hamiltonian-Rubin}
	\eeq
with $b=L,R$. The total Hamiltonian for the network, two baths and their couplings is $H_{T} = H + H_{L} + H_{R} + H_{c1}$. We write the equations of motion (EOM) for the network variables as: 
	\bea
	m_0 \ddot{y}_{\alpha,n} &=& - k_p y_{\alpha,n} - k \Big[(2 y_{\alpha,n} - y_{\alpha, n-1} - y_{\alpha,n+1}) \cr
	&+& \lambda (2 y_{\beta,n}- y_{\beta,n+1} - y_{\beta,n-1}) \Big] + \gamma_{L} y_{L,1}\delta_{n,1}\cr 
	&+& \gamma_{R} y_{R,1}\delta_{n,N},
	\label{e:EOM-for-system-variables-single-bath-Rubin}
	\eea
        for $\alpha \neq \beta, \alpha,\beta =1,2, n=1,2,\dots N$ and $y_{\alpha,0} = y_{\alpha, N+1} = 0$. 
The EOM for left and right bath variables are
	\bea
	m_0 \ddot{y}_{L,1} &=& -k_0 (2 y_{L,1} - y_{L,2}) + \gamma_L (y_{1,1} + y_{2,1}), \quad  \cr
        m_0 \ddot{y}_{R,1} &=& -k_0 (2 y_{R,1} - y_{R,2}) + \gamma_R (y_{1,N} + y_{2,N}), \quad  \cr	
	m_0 \ddot{y}_{b,n} &=& -k_0 (2 y_{b,n} - y_{b,n -1} - y_{b,n +1}),
	\label{e:EOM-for-left-bath-variables-single-bath-Rubin}
	\eea
for $b=L,R, n = 2,3, \cdots , N_b$. In order to solve the EOM for the bath variables, we rewrite the bath Hamiltonians in Eq.~\ref{e:bath-Hamiltonian-Rubin} in terms of the normal modes $Y_{b,r}$, by using the transformations, $y_{b,n} = \sum_{r=1}^{N_b}U_{b,nr} Y_{b,r}$ and $p_{b,n} = \sum_{r=1}^{N_b} U_{b,nr} P_{b,r}$, where $U_{b,nr} = (\sqrt{2/(N_b+1)}) \sin\left[ \pi nr/(N_b+1)\right]$. Thus, we have
	\beq
	H_{b} = \sum_{r=1}^{N_b} \frac{P_{b,r}^2}{2 m_0} + \frac{1}{2} m_0 \Omega_{b,r}^2 Y_{b,r}^2,
	\label{e:Bath-Hamiltonian-Rubin-Normal-mode}
	\eeq
for $b=L, R$, where
	\beq
	\Omega_{b,r}^2 = \Omega_{r}^2 = \frac{4 k_0}{m_0} \sin^2 \left( \frac{\pi r}{2(N_b+1)} \right).
	\label{e:bath-normal-frequencies}
	\eeq
In the last line, we assume $N_L=N_R$. Using these normal modes, we rewrite the coupling Hamiltonian in Eq.~\ref{e:system-bath-coupling-single-bath-Rubin} as: 
	\bea
	H_{c1}& =& - \Big[ \gamma_L (y_{1,1} + y_{2,1}) \sum_{r = 1}^{N_L} C_{L,r} Y_{L,r} \cr
	&+&\gamma_R (y_{1,N} + y_{2,N}) \sum_{r = 1}^{N_R} C_{R,r} Y_{R,r} \Big]. 
	\label{e:System-Bath-coupling-single-bath-Rubin-Normal-modes}
	\eea	
Here, $C_{b,r} = U_{b,1r}$. In terms of the aforementioned normal modes, we rewrite the EOM in Eq.~\ref{e:EOM-for-system-variables-single-bath-Rubin} as:
	\bea
	m_0 \ddot{y}_{\alpha,n} &=& - k_p y_{\alpha,n} - k \Big[(2 y_{\alpha,n} - y_{\alpha, n-1} - y_{\alpha,n+1}) \cr
	&+& \lambda (2 y_{\beta,n}- y_{\beta,n+1} - y_{\beta,n-1}) \Big] \cr &+& \gamma_{L} \sum_{r=1}^{N_L} C_{L,r} Y_{L,r}\delta_{n,1} \cr
	&+&\gamma_{R}\sum_{r=1}^{N_R} C_{R,r} Y_{R,r} \delta_{n,N}.
	\label{e:EOM-system-variables-single-bath-Rubin-Normal-mode}
	\eea
The EOM for the normal modes of baths follow from the Hamiltonians in Eqs.~\ref{e:Bath-Hamiltonian-Rubin-Normal-mode} and \ref{e:System-Bath-coupling-single-bath-Rubin-Normal-modes}:
	\bea
	m_0 \ddot{Y}_{L,r} &=& -m_0 \Omega_{r}^2 Y_{L,r} + \gamma_{L} C_{L,r} (y_{1,1} + y_{2,1}), \cr
	m_0 \ddot{Y}_{R,r} &=& -m_0 \Omega_{r}^2 Y_{R,r} + \gamma_{R} C_{R,r} (y_{1,N} + y_{2,N}). \quad
	\eea
We solve the above equations for the baths' modes at time $t>t_0$ with some initial conditions for $b=L,R$:
	\bea
	Y_{b,r}(t) &=& f^{+}_{r}(t - t_0) Y_{b,r}(t_0)+ g^{+}_{r}(t - t_0) \dot{Y}_{b,r}(t_0) \cr
	&+& \int_{t_0}^{t} dt' \Big[g^{+}_{r}(t - t') \frac{\gamma_b C_{b,r} \sqrt{2}}{m_0}\cr
	&&\times(y_{s,1}(t')\delta_{b,L}+y_{s,N}(t')\delta_{b,R}) \Big],
	\label{e:Formal-solution-left-bath-single-bath-Rubin}
	\eea
where $f^{+}_{r}(t)=\cos(\Omega_{r}t)\theta(t),g^{+}_{r}(t)=\sin(\Omega_{r}t)\theta(t)/\Omega_{r}$, and the symmetric modes $y_{s,n} = (y_{1,n} + y_{2,n})/\sqrt{2}$. Here, $\theta(t)$ is the Heaviside function. We can further write the EOM for the network variables in Eq.~\ref{e:EOM-system-variables-single-bath-Rubin-Normal-mode} at the boundaries in terms of the symmetric and anti-symmetric modes ($y_{a,n}  = (y_{1,n} - y_{2,n})/\sqrt{2}$) as: 
	\bea
	m_0 \ddot{y}_{s,1} &=&-k_p y_{s,1} - k \left[(2 y_{s,1} -y_{s,2}) + \lambda (2 y_{s,1} - y_{s,2}) \right] \cr
	&+& \sqrt{2} \gamma_L \sum_{r =1}^{N_L} C_{L,r} Y_{L,r}, \cr
	m_0 \ddot{y}_{s,N} &=& -k_p y_{s,N} - k \Big[(2 y_{s,N} - y_{s,N-1})  \cr 
	&+& \lambda (2 y_{s,N} - y_{s,N-1}) \Big] + \sqrt{2} \gamma_R \sum_{r=1}^{N_R} C_{R,r} Y_{R,r}, \cr
	m_0 \ddot{y}_{a,1} &=& -k_p y_{a,1} - k\left[(2 y_{a,1} - y_{a,2}) - \lambda (2 y_{a,1} - y_{a,2})\right], \cr
	m_0 \ddot{y}_{a,N} &=& -k_p y_{a,N} - k\Big[(2 y_{a,N} - y_{a,N-1})  \cr
	&-& \lambda (2 y_{a,N} - y_{a,N-1})\Big].  
	\label{e:EOM-sym-and-anti-symmetric}
	\eea
        We notice that the EOM of the anti-symmetric modes do not depend on the bath variables. Nevertheless, they still diverge in time in the inverted oscillator (unstable) regime. Plugging the formal solutions of the left bath variables in Eq.~\ref{e:Formal-solution-left-bath-single-bath-Rubin} in the EOM for the symmetric modes of the network, we get
	\bea
	m_0 \ddot{y}_{s,1} &=& -k_p y_{s,1} - k(2 y_{s,1} - y_{s,2}  + \lambda (2 y_{s,1} - y_{s,2})) \cr
	&+& \int_{t_0}^{t} dt' \Sigma_L^{+}(t-t') 2 y_{s,1}(t') + \sqrt{2} \eta_L(t).
	\label{e:EOM-symmetric-mode-single-bath-Rubin-left}
	\eea
Here, we identify the operator $\eta_L(t)$ as a noise from the left bath due to its dependence on the initial condition of the bath variables, and $\Sigma_L^{+}(t)$ as a self-energy arising from the coupling of the network to the left bath: 	
	\bea
	\eta_L(t)&=&\gamma_L \sum_{r=1}^{N_L} C_{L,r} \Big(f^{+}_{r}(t-t_0)Y_{L,r}(t_0) \cr
	&+&g^{+}_{r}(t - t_0)\dot{Y}_{L_{\beta}}(t_0) \Big), 
	\label{e:Noise-Rubin-bath} \\
	\Sigma_L^{+}(t)&=& \sum_{r=1}^{N_L} \frac{\gamma_L^2 C_{L,r}^2}{m_0}g^{+}_{r}(t).
	\label{e:Memory-Kernal-Rubin-bath}
	\eea  
Similarly, using the noise and self-energy for the right bath, we obtain:  
	\bea
	m_0 \ddot{y}_{s,N} &=& -k_p y_{s,N} - k \Big[(2 y_{s,N} - y_{s,N-1})\cr
	&+&  \lambda (2 y_{s,N} - y_{s,N-1}) \Big] \cr
	&+& \int_{t_0}^{t} dt' \Sigma_R^{+}(t-t') 2 y_{s,N}(t') + \sqrt{2} \eta_R(t).  \qquad
	\label{e:EOM-symmetric-mode-single-bath-Rubin-right}
	\eea
The above equations along with those for the anti-symmetric modes in Eq.~\ref{e:EOM-sym-and-anti-symmetric} are the EOM in Eqs.~\ref{e:eom1} and \ref{e:eom2} in terms of the original system variables after letting $t_0 \to -\infty$. After Fourier transformation to the frequency domain, the EOM in Eqs.~\ref{e:EOM-symmetric-mode-single-bath-Rubin-left}, \ref{e:EOM-symmetric-mode-single-bath-Rubin-right} and \ref{e:EOM-sym-and-anti-symmetric} transform into Eqs.~\ref{e:fm1}, \ref{e:fm2}, \ref{e:fm3}, and \ref{e:fm4}.

The self-energy due to bath coupling can be computed as
	\bea
	\tilde{\Sigma}^{+}_b(\omega) &=& \int_0^{\infty} \sum_{r} \frac{\gamma_b^2 C_{b,r}^2}{m_0}  \frac{\sin(\Omega_{r} t)}{\Omega_{r}} e^{i \omega t} dt \cr
&=& \lim_{\epsilon \to 0} \int_0^{\infty} \sum_{r} \frac{\gamma_b^2 C_{b,r}^2}{m_0 \Omega_{r}} \frac{e^{i \Omega_{r} t} - e^{- i \Omega_{r} t}}{2 i} e^{i \omega t - \epsilon t} dt \cr
&=&\frac{\gamma_{b}^2}{k_0} \Big[ \left( 1 - \frac{m_0 \omega^2}{2k_0} \right) \cr
&+&i \left( \frac{\sqrt{m_0}\omega}{\sqrt{k_0}} \left( 1 - \frac{m_0 \omega^2}{4 k_0} \right)^{1/2} \right) \Big], 
	\label{e:Memory-Kernel-Rubin-Fourier}
	\eea
for $b=L,R$. The properties of noises are determined by the equilibrium conditions of the isolated baths. Thus, $\left \bra \eta_b(t)\right \ket =0,$ and the noise correlation is given by the fluctuation-dissipation relation as 
	\beq
	\left \bra \tl{\eta}_b(\omega) \tl{\eta}_{b'}(\omega') \right \ket =  \tl{\Gamma}_b(\omega) \frac{\hbar}{\pi}(1 + f_b) \delta_{b,b'}\delta(\omega + \omega'),
	\label{e:FD-theorem}
	\eeq
where $\tilde{\Gamma}_b(\omega) = {\rm Im} (\tilde{\Sigma}_b^{+}(\omega))$, and the equilibrium phonon distribution functions $f_b \equiv f(\omega, T_{b}) = 1/ (e^{\hbar \omega/ k_B T_{b}}-1)$, for $b,b'=L,R$. Here, $\left \bra \dots \right \ket$ denotes an expectation over isolated baths' equilibrium distribution.  

\subsection{Linear response heat current}	

We define the heat current at the left boundary of the network as the rate of work done by the left bath on the network, which is given by the expression (see Eq.~\ref{e:heat-current-J1}):
	\bea
	J_{\rm I}& =& -\Big[ \dot{y}_{1,1} \Big(\int_{-\infty}^t dt'\Sigma_L^{+}(t-t') \big( y_{1,1}(t')+y_{2,1}(t')\big)+\eta_L \Big) \cr
	&+& \dot{y}_{2,1} \Big(\int_{-\infty}^t dt'\Sigma_L^{+}(t-t')\big(y_{1,1}(t') + y_{2,1}(t')\big)+ \eta_L \Big) \Big].\nn\\
	\label{e:current-single-bath-Rubin-time-domain}
	\eea	
We express the noise averaged heat current in terms of the symmetric modes in the frequency domain as:
	\bea
	\bra J_{\rm I} \ket  &=& -\int_{-\infty}^{\infty} d\omega \int_{-\infty}^{\infty} d\omega' e^{-i(\omega + \omega')t} i \omega \Big 	\bra \tl{y}_{s,1}(\omega) \cr 
	&&\left( \sqrt{2} \tl{\eta}_L(\omega') + 2 \tilde{\Sigma}_L^{+}(\omega') \tilde{y}_{s,1}(\omega') \right)\Big \ket. 
	\label{e:current-single-bath-Rubin-frequency-domain}
	\eea
Here, 
	\bea
	\tl{y}_{s,1}(\omega)=[G_s^+(\omega)]_{1,1} \sqrt{2} \tl{\eta}_L + [G_s^+(\omega)]_{1,N}  \sqrt{2} \tilde{\eta}_R,
	\eea
where, $G_s^{+}(\omega)$ is the inverse of the tri-diagonal matrix $Z$:
	\bea
	Z = \begin{bmatrix}
	z_{L +} & -\lambda_{+}& 0 & \cdots & & 0 \\
	-\lambda_{+}  & z_{+}  & -\lambda_{+} & \ddots & &\vdots \\
	\vdots & \ddots & \ddots & \ddots &  \ddots &  \vdots \\
	\vdots &  & \ddots & \ddots & \ddots & \vdots\\
	\vdots  & & & -\lambda_{+} & z_{+} 	&  -\lambda_{+}\\
	0 &  \cdots & \cdots & & 	-\lambda_{+} & z_{R +}  \\
	\end{bmatrix}. \quad
	\label{e:symmetric-Z-matrix}
	\eea	
Here, $z_{b+} = -m_0 \omega^2 + k_p + 2 \lambda_{+} - 2 \tilde{\Sigma}_{b}^{+}(\omega)$, $z_{+} =  -m_0 \omega^2 + k_p + 2\lambda_{+}$, and $\lambda_{+} = k(1 + \lambda)$ for $b=L,R$. Thus, we find from Eq.~\ref{e:current-single-bath-Rubin-frequency-domain}:
	\bea
	\bra J_{\rm I} \ket &=& -\int_{-\infty}^{\infty} d\omega \int_{-\infty}^{\infty} d\omega' e^{-i(\omega + \omega')t} 2i \omega \times \cr
	&& \Big \bra \big( [G_s^+(\omega) ]_{1,1}\tl{\eta}_L(\omega)+[G_s^+(\omega)]_{1,N}\tl{\eta}_R(\omega)\big)\cr
	&& \Big(\tl{\eta}_L(\omega') + 2 \tl{\Sigma}^+_L(\omega')\big([G_s^+(\omega')]_{1,1}\tl{\eta}_L(\omega') \cr
	&&+ [G_s^+ (\omega')]_{1,N} \tl{\eta}_R(\omega')\big)\Big) \Big \ket.
	\label{e:heatcur-RG-J1-fm}
	\eea
The contribution of the right bath to $\bra J_{\rm I} \ket$ in Eq.~\ref{e:heatcur-RG-J1-fm} can be obtained using the fluctuation-dissipation relation for the noise averaging (see Eq.~\ref{e:FD-theorem})  as:
	\bea
	\bra J_{\rm I}^R \ket &=& -\int_{-\infty}^{\infty} d\omega \, \Big[ 4 |[G_s^+(\omega)]_{1,N}|^2 \tl{\Gamma}_L(\omega)\tl{\Gamma}_R(\omega)\cr
	&&\frac{\hbar \omega}{\pi}(1 + f_R) \Big].
	\eea
Similarly, we can find the contribution from the left bath to $\bra J_{\rm I} \ket$ in Eq.~\ref{e:heatcur-RG-J1-fm}. We choose $\tl{\Gamma}_L(\omega)=\tl{\Gamma}_R(\omega)$ (which are nonzero only when $|\omega| < 2\sqrt{k_0}$) by setting $\gamma_L = \gamma_R = \gamma$ to get 
	\bea
	\bra J_{\rm I} \ket &=& \int_{-2\sqrt{k_0}}^{2 \sqrt{k_0}} d\omega \, \left[ 4 |[G_s^+(\omega)]_{1,N}|^2 \frac{m_0 \omega^2 \gamma^4}{k_0^3} \right. \cr
	&&\left. \left( 1 - \frac{m_0 \omega^2}{4k_0} \right)  \frac{\hbar \omega}{\pi}(f_L - f_R) \right].
	\label{e:quant-current-single-bath-Rubin}
	\eea
In the linear response regime, after expanding the Bose functions, we get the quantum heat current (see Eq.~\ref{e:heatcur-RG-J1-lin-res})
	\bea
	\bra J_{\rm I} \ket &=& \frac{k_B \Delta T}{\pi} \int_{-2\sqrt{k_0}}^{2 \sqrt{k_0}} d\omega \, \left[ 4 |[G_s^+(\omega)]_{1,N}|^2 \frac{m_0 \omega^2 \gamma^4}{k_0^3} \right. \cr
	&&\left.\left( 1 - \frac{m_0 \omega^2}{4k_0} \right)  \left( \frac{\hbar \omega}{2 k_B T} \right)^2 {\rm csch}^2 \left( \frac{\hbar \omega}{2 k_B T} \right) \right]. \quad \qquad
	\label{e:quant-current-single-bath-Rubin-Linear-Response}
	\eea
In the high-temperature limit, we obtain the classical heat current in Eq.~\ref{e:heatcur-RG-J1-clas}. Next, we find an analytical expression for this classical heat current in the thermodynamic limit. 

\subsection{Thermodynamic limit and an analytic expression for classical heat current}	

To find an analytical expression for the classical current in Eq.~\ref{e:heatcur-RG-J1-clas}, we need to manipulate the properties of the matrix $Z=\lambda_{+}Z_s$ in the large $N$ limit. Given $N$ number of lattice sites in the network, $Z_s$ is an $N \times N$ tridiagonal matrix of the form given in Sec.~\ref{s:Single-bath}, and we define $\tilde{a}=(-m_0 \omega^2 + k_p)/\lambda_{+}+2$ and $\epsilon_{L,R} = 2\tl{\Sigma}^+_{L,R}/\lambda_{+}$. We recall that the expression for quantum heat current in Eq.~\ref{e:quant-current-single-bath-Rubin} involves the matrix element $[G_s^+]_{1,N}$, which is given by the inverse of $Z$ matrix: $G_s^+ =Z^{-1} = Z_s^{-1}/\lambda_{+}$. From this, we get:
	\beq
	|[G_s^{+}]_{1,N}|^2 = \frac{1}{\lambda_{+}^2 |\det [Z_s]_N|^2}.
	\eeq
The determinant of a tridiagonal matrix can be obtained through a recursion relation. If all the diagonal entries are $\tilde{a}$, then the determinant of an $N\times N$ matrix is given by the recursion relation:
	\beq
	\det([Z_s]_N) = \tilde{a}\det([Z_s]_{N-1}) - \det([Z_s]_{N-2}).
	\eeq
We may simplify the expression for the determinant by assuming $\tilde{a} = 2 \cos q$. To simplify this calculation, we further assume $\omega^2 = 4k_0 \sin^2(q/2)$, which restricts the following parameters as: $m_0 = 1, k_p = 0$ and $\lambda_{+} = k_0$. With these simplifications, the determinant takes the form 
	\beq
	\det [Z_s]_N = \Delta_N = \frac{(A(q) \sin(N q) + B(q) \cos(N q))}{\sin q}.
	\label{e:determinant-of-tri-diag-matrix}
	\eeq
Here, 
	\bea
	A(q) &=& -(\epsilon_L + \epsilon_R) + (1 + \epsilon_L \epsilon_R) \cos(q), \quad \cr B(q) &=& (1 - \epsilon_L \epsilon_R) \sin(q).
	\label{e:A-and-B-functions-for-determinant}
	\eea 
From this, we find that 
	\beq
	|\Delta_N |^2 = \frac{(|A|^2 + |B|^2)(1 + r \sin(2Nq + \phi))}{2 \sin^2 q},	
	\label{e:det-square-of-tri-diag-matrix}
	\eeq
with $r\cos \phi = (A B^{*} + A^{*} B)/(|A|^2 + |B|^2)$, and $r\sin \phi = (|B|^2 - |A|^2)/(|A|^2 + |B|^2)$.

By substituting $\omega^2 = 4k_0 \sin^2(q/2)$, we get the expression for the classical heat current:
	\bea
	\bra J_{\rm I}^{cl} \ket  &=& \frac{k_B \Delta T}{\pi} \int_{-\pi}^{\pi} dq \, \Big[ |[G_s^+(q)]_{1,N}|^2 \cr
	&&\times \sqrt{k_0} \cos \left(\frac{q}{2} \right)\left(\frac{2 \gamma^2}{k_0} \right)^2 \sin^2(q) \Big].
	\label{e:class-current-single-bath-each-boundary-Rubin}
	\eea
For the Rubin baths, we can write the self-energy as $\tl{\Sigma}_b^{+}(q) = (\gamma^2/k_0)e^{iq}$ [see Eq.~\ref{e:Memory-Kernel-Rubin-Fourier} along with $\omega^2 = 4k_0 \sin^2(q/2)$]. Thus, we can express 
	\bea
	A(q)&=&\frac{-4 \gamma^2}{k_0^2} \cos q + \left( 1 + \frac{4 \gamma^4}{k_0^4} \cos 2q \right) \cos q \cr
	&+&i \left( \frac{-4 \gamma^2}{k_0^2} \sin q + \left(\frac{4 \gamma^4}{k_0^4} \sin 2q\right) \cos q \right), \qquad
	\eea
and 
	\beq
	B(q) = \left( 1 - \frac{4 \gamma^4}{k_0^4} \cos 2q\right) \sin q - i \frac{4 \gamma^4}{k_0^4} \sin 2q \sin q.
	\eeq
Employing the above expressions, we rewrite Eq.~\ref{e:class-current-single-bath-each-boundary-Rubin} as 
	\bea
	\bra J_{\rm I}^{cl} \ket &=& \frac{k_B \Delta T}{\pi} \int_{-\pi}^{\pi}\frac{dq \sqrt{k_0}}{k_0^2|\Delta_N|^2}  \cos \left(\frac{q}{2} \right)\left(\frac{2 \gamma^2}{k_0} \right)^2 \sin^2(q) \cr
	&=& \frac{k_B\Delta T 16 \gamma^4}{\pi k_0^3 \sqrt{k_0}} \int_{0}^{\pi}  dq \,\Big[\frac{\cos \left(\frac{q}{2} \right) \sin^4(q) }{(|A|^2+|B|^2)} \times \cr
	&&  \frac{1}{(1 + r \sin(2 N q + \phi))} \Big].
	\label{e:heatcur-RG-J1-clas-in-q}
	\eea
In the large $N$ limit, we can simplify the above integral using the identity \cite{Roy_2008},
	\beq
	\lim_{N \to \infty} \int_{0}^{\pi} dq \frac{g_1(q)}{1 + g_2(q) \sin(N q)} = \int_{0}^{\pi} dq \frac{g_1(q)}{(1- g_2(q)^2)^{1/2}}, 			\
	\label{e:large-N-limit-identity}
	\eeq	
where, we identify by comparing the above equation with Eq.~\ref{e:heatcur-RG-J1-clas-in-q}, $g_2(q) = r$ and
	\beq
	g_1(q) = \frac{\cos(q/2) \sin^4(q)}{(|A|^2 + |B|^2)}.
	\eeq	
Using $r$ (see below of Eq.~\ref{e:det-square-of-tri-diag-matrix} for definitions) and the re-definitions of $A = a_1 + i a_2$ and $B = b_1 + i b_2$, we have
	\beq
	(1 - g_2(q))^{1/2} = \frac{2 (a_1 b_2 - a_2 b_1)}{|A|^2 + |B|^2}.
	\label{e:large-N-limit-integrand-simplification-relation}
	\eeq 
Thus, in the large $N$ limit, the integrand in Eq.~\ref{e:heatcur-RG-J1-clas-in-q} simplifies to 	\beq
	\int_{0}^{\pi} dq \frac{g_1(q)}{(1- g_2(q))^{1/2}} = \int_{0}^{\pi} dq  \frac{\cos(q/2) \sin^4 q}{2(|a_1 b_2 - a_2 b_1|)}.
	\eeq
We further derive using $A(q)$ and $B(q)$  (see Eq.~\ref{e:A-and-B-functions-for-determinant}), 
	\bea
&&	2(a_1 b_2 - a_2 b_1) = \frac{8 \gamma^2 \sin^2 q (\Upsilon - \Omega \cos(2 q))}{k_0^6}, \nn\\
&&	\Upsilon = k_0^4 - 2 k_0^2 \gamma^2 + 4 \gamma^4 \quad \text{and} \quad \Omega = 2 k_0^2 \gamma^2.
	\eea 
We thus write the classical heat current as 
	\bea
	\bra J_{\rm I}^{cl} \ket & =& \frac{k_B \Delta T}{\pi} \frac{2 \gamma^2 k_0^3}{ \sqrt{k_0}} \int_{0}^{\pi} \frac{ \cos(q/2) \sin^2 q \,  dq }{\Upsilon - \Omega \cos(2 q)}\nn\\ &=& \frac{k_B \Delta T \Omega \sqrt{k_0}}{\pi} \int_{-1}^{1} dx \frac{\sqrt{\frac{1+x}{2}}\sqrt{1-x^2}}{\Upsilon - \Omega (2 x^2 -1)}, \quad
	\eea
	where we use the substitution $x = \cos q$ in the last line. We find Eq.~\ref{e:heatcur-RG-J1-clas-analy-exp} by performing the above integral. Numerical integration of the integral in Eq.~\ref{e:heatcur-RG-J1-clas-in-q} (for large $N$ limit) agrees with the value that obtained directly by numerically inverting the matrix for sufficiently large $N$.

\subsection{General case: Analytical expression for classical current} 

We may now consider a more general case where we {\it do not} restrict the parameters as $m_0 = 1, k_p = 0$ and $k_0 = \lambda_{+}$. We start with the same substitution $\tilde{a} = 2 \cos q$. However, now $\omega^2 \neq 4 k_0 \sin^2(q/2)$, since $m_0 \neq 1, k_p \neq 0$ and $k_0 \neq \lambda_{+}$. We employ the expression of $\tl{\Sigma}^{+}(\omega)$ in Eq.~\ref{e:Memory-Kernel-Rubin-Fourier}, and the definitions of $A$ and $B$ in Eq.~\ref{e:A-and-B-functions-for-determinant} to obtain
\begin{widetext}
	\bea
	2(a_1 b_2 - a_2 b_1) &=&  \frac{4 \gamma^2 \sqrt{(k_p + 4 \lambda_{+} \sin^2(q/2))(4 k_0  - k_p - 4 \lambda_{+} \sin^2(q/2))}}{k_0^4}  \cr
	&&\times \frac{\big(4 \gamma^4 + k_0^2 \lambda_{+}^2 - 2 \lambda_{+} \gamma^2 (\cos q) (2 k_0 - k_p - 4 \lambda_{+} \sin^2(q/2))\big) \sin q}{\lambda_{+}^3}. 
	\label{e:part-of-integral-large-N}
	\eea
\end{widetext}
For a nonzero $k_p$, we express the linear response classical heat current as: 
	\bea
	\bra J_{\rm I}^{cl} \ket &=& \frac{k_B \Delta T}{\pi} \int_{-\omega_{1+}}^{\omega_{1+}} d\omega |[G_s^+(\omega)]_{1,N}|^2 \cr
	&&\times \frac{4 m_0 \omega^2 \gamma^4}{k_0^3}  \left( 1- \frac{m_0 \omega^2}{4 k_0} \right),
	\eea
where $\omega_{1+} = \sqrt{(k_p + 4 \lambda_{+})/m_0}$. Substituting $\omega^2 = (k_p + 4 \lambda_{+} \sin^2(q/2))/m_0$ in the above expression, we get
	\bea
	\bra J_{\rm I}^{cl} \ket &=& \frac{k_B \Delta T}{\pi} \int_{-\pi}^{\pi} dq\Big[ \left| \frac{d \omega}{dq} \right| |[G_s^+(\omega)]_{1,N}|^2 \cr
	&&\times  \frac{4 m_0 \omega^2 \gamma^4}{k_0^3}\left( 1- \frac{m_0 \omega^2}{4 k_0} \right) \Big].   	
	\eea
where, $|d\omega/ dq| = (\lambda_{+}\sin q)/ (m_0 \omega)$. Applying the expression for $\Delta_N$ in the large $N$ limit (as in Eq.~\ref{e:large-N-limit-identity}), we re-express the above current as
\begin{widetext}	
	\beq
\bra J_{\rm I}^{cl} \ket = \frac{k_B \Delta T}{\pi} \int_{0}^{\pi} dq \frac{(\sin^2 q) \gamma^2 \lambda_{+}^2 \sqrt{4 k_0  - k_p - 4 \lambda_{+} \sin^2(q/2)}}{ \sqrt{m_0}\big( 4 \gamma^4 + k_0^2 \lambda_{+}^2 - 2 \gamma^2 \lambda_{+} (\cos q) (2 k_0 - k_p - 4 \lambda_{+} \sin^2(q/2)) \big)},
	\eeq
where we use Eq.~\ref{e:part-of-integral-large-N} along with the form of Eq.~\ref{e:large-N-limit-integrand-simplification-relation}. We substitute $x = \cos q$ to get 
	\beq
	\bra J_{\rm I}^{cl} \ket  = \frac{k_B \Delta T}{\pi \sqrt{m_0}} \int_{-1}^{1}  \frac{dx \, \gamma^2 \lambda_{+}^2 \sqrt{1 - x^2} \sqrt{4 k_0  - k_p - 2 \lambda_{+}(1-x)}}{4 \gamma^4 + k_0^2 \lambda_{+}^2 - 2 \gamma^2 \lambda_{+} x (2 k_0 - k_p -2 \lambda_{+} (1-x))}. 
	\eeq
The above definite integral can be expressed in terms of complete and incomplete elliptic integrals of first ($K$ and $F$), second ($E$) and third kind ($\Pi$). Thus, we obtain
	\bea
	\bra J_{\rm I}^{cl} \ket &=& \frac{k_B \Delta T}{4 \pi \sqrt{m_0} \sqrt{k_p - 4 k_0 + 4 \lambda_{+}}} \left[ -2(k_p + 4 				\lambda_{+} - 4 k_0) (E(\phi, \zeta) + E(\zeta))  + 4(\lambda_{+} - k_0 ) (F(\phi, \zeta) + K(\zeta)) \right. \cr
	&&\left. + (k_p + 2 \lambda_{+} -2 k_0 ) \left(\Pi(\xi_{-}, \phi, \zeta) + \Pi(\xi_{+}, \phi, \zeta) + \Pi(\xi_{-}, \zeta) + \Pi(\xi_{+}, 		\zeta) \right) \right. \cr		
	&&\left. + \frac{\gamma^2 \left(4 k_0^2 - 4 k_0 (k_p + 2 \lambda_{+}) + k_p^2 + 4 k_p \lambda_{+} - 4 \lambda_{+}^2 \right) 		+ 2 k_0^2 \lambda_{+}^2  + 8 \gamma^4}{\sqrt{4 \gamma^2 k_0^2 \lambda_{+}^2  + \gamma^4 (-2 k_0  + k_p + 2 			\lambda_{+})^2 + 16 \gamma^6}} \right. \cr
	&&\times \left. \left (\Pi(\xi_{-}, \phi, \zeta) - \Pi(\xi_{+}, \phi, \zeta) + \Pi(\xi_{-}, \zeta) - \Pi(\xi_{+}, \zeta) \right) \right],	
	\eea
where
	\bea
	\phi &=& i \sinh^{-1} \left(\sqrt{\frac{4 k_0  - k_p -4 \lambda_{+}}{k_p - 4 k_0}}\right), \quad \zeta = \frac{k_p - 4 k_0}{k_p + 4 		\lambda_{+} - 4 k_0}, \nn\\
	\xi_{\pm} &=&\frac{2 \gamma ^2 (k_p- 4 k_0)}{\gamma^2 (-6 k_0 + k_p + 2 \lambda_{+}) \pm  \sqrt{2 \Omega \lambda_{+}^2 	+ \gamma^4 (-2 k_0 + k_p + 2 \lambda_{+})^2 + 16 \gamma^6}}.  
	\eea
\end{widetext}

\subsection{Temperature dependence of low-temperature $\bra J_{\rm I}\ket$}	

Here, we determine the temperature dependence of the low-temperature heat current $\bra J_{\rm I}\ket$. The quantum current $\bra J_{\rm I}\ket$ in the linear response regime reads from Eq.~\ref{e:quant-current-single-bath-Rubin-Linear-Response} as:
	\bea
	&&\bra J_{\rm I} \ket  = \frac{4k_B \Delta T}{\pi} \int_{-\pi}^{\pi} dq \left[\left| \frac{d\omega}{dq}\right| |[G_s^{+}(q)]_{1,N}|^2 \frac{m_0 \omega^2 \gamma^4}{k_0^3}  \times \right.\cr
	&&~~ \left.\left( 1 - \frac{m_0 \omega^2}{4 k_0}\right) \left( \frac{\hbar \omega}{2 k_B T}\right)^2 {\rm csch}^2 \left( \frac{\hbar \omega}{2 k_B T} \right) \right],
	\label{e:heatcur-RG-J1-quant}
	\eea
where $\omega^2 = (k_p + 4 \lambda_{+} \sin^2(q/2))/m_0$. Since the low-frequency or low-wavevector modes mostly contribute in heat conduction at low temperature, we determine the temperature dependence of low-temperature $\bra J_{\rm I}\ket$ by studying the above integrand in the limit of small $\omega$ or $q$. We rewrite the integrand in Eq.~\ref{e:heatcur-RG-J1-quant} in the large $N$ limit in terms of the system parameters as:
\begin{widetext}
	\bea
	\bra J_{\rm I} \ket = \frac{\hbar^2 \Delta T }{4 \pi k_B \sqrt{m_0}T^2} \int_{0}^{\pi} dq \, \Big[\frac{\gamma^2 \lambda_{+}^2 \sin^2(q) \sqrt{4 k_0  - m_0 \omega^2}}{\Lambda - 2 \gamma^2 \lambda_{+} \cos (q)(2 k_0 - m_0 \omega^2)} \times\omega^2 {\rm csch}^2\left( \frac{\hbar \omega}{2 k_B T}\right)\Big], 
	\label{e:heatcur-RG-J1-quant-sys-para}
	\eea
where, $\Lambda = (4 \gamma^4 + k_0^2 \lambda_{+}^2)$. We take the limit of small $q$ for $k_p \neq 0$ to find:
	\beq
	\bra J_{\rm I} \ket \approx \frac{\hbar^2 \Delta T \:e^{-\frac{ \hbar \omega_0}{k_B T}}}{\pi k_B \sqrt{m_0} T^2} \int_{0}^{\pi} dq \Big[ \frac{ \gamma^2 \lambda_{+}^2 q^2 \sqrt{4 k_0 - k_p - \lambda_{+} q^2}}{ ( \Lambda  -  2 \gamma^2 \lambda_{+} (1 - q^2/2) (2 k_0  - k_p - \lambda_{+} q^2))} \left( \frac{k_p}{m_0} + \frac{\lambda_{+} q^2}{m_0}\right) e^{- \frac{\hbar \lambda_{+} q^2}{2 k_B T m_0 \omega_0}} \Big],
	\eeq
\end{widetext}	
where we approximate $\omega^2 \approx (k_p + \lambda_{+} q^2)/m_0$ and ${\rm csch}^2\left[ \hbar \omega/ (2 k_B T) \right] \approx 4 e^{-\hbar \omega_0/(k_B T)} e^{- \hbar \lambda_{+} q^2/(2 k_B T m_0 \omega_0)}$. By substituting $x = (\hbar \lambda_{+}q^2)/(2 k_B T m_0 \omega_0)$, we obtain the leading-order $T$-dependence of $\bra J_{\rm I} \ket$ for $k_p \neq 0$ as:
	\beq
	\bra J_{\rm I} \ket \sim \frac{e^{\frac{-\hbar \omega_0}{k_B T}}}{\sqrt{T}} \quad  \text{with} \quad \omega_0 = \sqrt{\frac{k_p}{m_0}}.
	\eeq 
For $k_p = 0$, we notice that $\omega \propto q$ for small $q$. We substitute $x = (\hbar \omega)/(2 k_B T)$ in Eq.~\ref{e:heatcur-RG-J1-quant-sys-para}. Moreover, the denominator of the integrand in Eq.~\ref{e:heatcur-RG-J1-quant-sys-para} can be re-expressed in the small $q$ limit as 
	\beq
	(2 \gamma^2 - k_0 \lambda_{+})^2 + (2 k_0 \lambda_{+} \gamma^2 + 2 \gamma^2 \lambda_{+}^2) q^2.
	\eeq 
Thus, analytically, we get $\bra J_{\rm I} \ket \sim T$, when $2\gamma^2 = k_0 \lambda_{+}$ and $\bra J_{\rm I} \ket \sim T^3$ when $2 \gamma^2 \neq  k_0 \lambda_{+}$. We verified these results numerically. 

\section{Two Rubin baths at each boundary}
\label{b:Two-baths-at-each-boundary-Rubin-Model}

In this appendix, we consider the case where two types of oscillators at the left and right end of the network are connected to two different Rubin baths. Thus, the network-bath coupling is given by the Hamiltonian:
	\bea
	H_{c2} &=& -\gamma_L^{(1)} y_{L,1}^{(1)} y_{1,1} -\gamma_L^{(2)} y_{L,1}^{(2)} y_{2,1} \cr
	&-&\gamma_R^{(1)} y_{R,1}^{(1)} y_{1,N} -\gamma_R^{(2)} y_{R,1}^{(2)} y_{2,N}.
	\eea
Here, $ y_{L, n}^{(1)/(2)}$ represent two different types of left bath variables (differentiated by the superscripts (1) and (2)), and $y_{R,n}^{(1)/(2)}$ represent those for the right baths. Moreover, $\gamma_L^{(1)/(2)}$ and $\gamma_R^{(1)/(2)}$ are the coupling strengths between the  left and right ends of the network and the corresponding baths, respectively. The total Hamiltonian for the network and four baths including the coupling is $H_T = H + H_{L}^{(1)} +  H_{L}^{(2)}+  H_{R}^{(1)} +  H_{R}^{(2)} + H_{c2}$. Here, $H_{b}^{(1)/(2)}$ (for $b = L, R$) are the Hamiltonians of two types of Rubin baths at each end, as defined in Eq.~\ref{e:bath-Hamiltonian-Rubin} of Appendix \ref{a:Single-bath-at-each-boundary-Rubin-Model}. At first, we write down the EOM for the network variables  in terms of the normal modes of the bath variables (see Eq.~\ref{e:Bath-Hamiltonian-Rubin-Normal-mode}) as:
	\bea
	m_0 \ddot{y}_{\alpha,n} &=& - k_p y_{\alpha,n} - k \Big[(2 y_{\alpha,n} - y_{\alpha, n-1} - y_{\alpha,n+1}) \cr
	&+& \lambda (2 y_{\beta,n}- y_{\beta,n+1} - y_{\beta,n-1}) \Big] \cr 
	&+&\gamma_{L}^{(1)} \sum_{r=1}^{N_L} C_{L,r} Y_{L,r}^{(1)}\delta_{n,1} \delta_{\alpha,1} \cr
	&+& \gamma_{R}^{(1)}\sum_{r=1}^{N_R} C_{R,r} Y_{R,r}^{(1)} \delta_{n,N} \delta_{\alpha, N}\cr
	&+&\gamma_{L}^{(2)} \sum_{r=1}^{N_L} C_{L,r} Y_{L,r}^{(2)}\delta_{n,1} \delta_{\alpha,2}\cr
	&+&\gamma_{R}^{(2)}\sum_{r=1}^{N_R} C_{R,r} Y_{R,r}^{(2)} \delta_{n,N}  \delta_{\alpha,2},
	\label{e:EOM-system-variable-two-diff-Rubin-baths}
	\eea
for $\alpha \neq \beta, \alpha,\beta =1,2, n=1,2,\dots N$ and $y_{\alpha,0} = y_{\alpha, N+1} = 0$.	Here,  $y_{b,n}^{(1)/(2)} = \sum_{r=1}^{N_b}U_{b,nr} Y_{b,r}^{(1)/(2)}$ and $C_{b,r} = U_{b,1r}$ for $b = L, R$ (see Appendix \ref{a:Single-bath-at-each-boundary-Rubin-Model} for details). Similarly, we obtain the  EOM for the bath variables: 
	\bea
	m_0 \ddot{Y}_{L,r}^{(1)/(2)} &=& -m_0 \Omega_{L, r}^2 Y_{L,r}^{(1)/(2)} + \gamma_{L}^{(1)/(2)} C_{L,r} y_{1/2,1}, \cr
	m_0 \ddot{Y}_{R,r}^{(1)/(2)} &=& -m_0 \Omega_{R, r}^2 Y_{R,r}^{(1)/(2)} + \gamma_{R}^{(1)/(2)} C_{R,r} y_{1/2,N}. \nn \\ 
	\eea
Here, $\Omega_{b, r}^2 = \Omega_{r}^2 = (4 k_0/m_0) \sin^2 \left[\pi r/(2(N_b+1)) \right]$ are the normal frequencies assuming $N_L = N_R$. Consequently, the formal solutions at $t > t_0$ are:
	\bea
	Y_{b,r}^{(1)/(2)}(t) &=& f_r^{+}(t - t_0)) Y_{b,r}^{(1)/(2)}(t_0) \cr
	&+& g_r^{+}(t- t_0) \dot{Y}_{b,r}^{(1)/(2)}(t_0) \cr
	&+& \int_{t_0}^{t} dt' \Big[g_r^{+}(t- t')  \frac{\gamma_b^{(1)/(2)} C_{b,r}}{m_0} \cr
	&&\times (y_{1/2,1}(t') \delta_{b,L} +  y_{1/2,N}(t') \delta_{b,R}) \Big],\quad
	\eea
where $f^{+}_{r}(t)=\cos(\Omega_{r}t)\theta(t),g^{+}_{r}(t)=\sin(\Omega_{r}t)\theta(t)/\Omega_{r}$. Substituting these formal solutions in Eq.~\ref{e:EOM-system-variable-two-diff-Rubin-baths} for the boundary variables imply Eqs.~\ref{e:eom3} and \ref{e:eom4} after letting $t_0 \to -\infty$, provided we identify the noises $\eta_{1/2,L}(t)$ and self-energies $\Sigma_{1/2,L}^{+}(t)$ as:
	\bea
	&&\eta_{1/2,L}(t) = \gamma_L^{(1)/(2)} \sum_{r =1}^{N_L} C_{L,r} \Big( f_r^{+}(t-t_0)   \cr
	&&~~ Y_{L,r}^{(1)/(2)}(t_0) + g_r^{+}(t-t_0) \dot{Y}_{L,r}^{(1)/(2)}(t_0) \Big), \quad
	\eea
and
	\beq
	\Sigma_{1/2,L}^{+}(t) = \sum_{r=1}^{N_L} \frac{\left(\gamma_L^{(1)/(2)}\right)^2 C_{L,r}^2}{m_0} g_r^{+}(t).
	\eeq 
Similar definitions hold for the noises $\eta_{1/2,R}(t)$  and self-energies $\Sigma_{1/2,R}^{+}(t)$ of the right baths. Now, we rewrite these equations (assuming same coupling strengths $\gamma_b^{(1)} = \gamma_b^{(2)}$, which implies $\Sigma_{1,b}^{+} = \Sigma_{2,b}^{+} = \Sigma_b^{+}$ for $b = L,R$) in terms of the symmetric and anti-symmetric variables as:
	\bea
	m_0 \ddot{y}_{s,1} &=& -k_p y_{s,1} - k[2 y_{s,1} - y_{s,2} + \lambda (2 y_{s,1} - y_{s,2})] \cr
	&+&\int_{t_0}^{t} dt' \Sigma_L^{+}(t-t') y_{s,1}(t')  +  \eta_{s,L},
	\label{e:EOM-sym-1-diff-bath-Rub}\\
	m_0 \ddot{y}_{s,N} &=& -k_p y_{s,N} - k\Big[2 y_{s,N} - y_{s,N-1} \cr 
	&+& \lambda (2 y_{s,N} - y_{s,N-1})\Big]  \cr
	&+& \int_{t_0}^{t} dt' \Sigma_R^{+}(t-t') y_{s,N}(t') + \eta_{s,R}.
	\label{e:EOM-sym-N-diff-bath-Rub}
	\eea
Here, $\eta_{s/a,L} = (\eta_{1,L}(t) \pm \eta_{2,L}(t))/\sqrt{2}$. Similarly, for the anti-symmetric variables, we get 
	\bea
	m_0 \ddot{y}_{a,1} &=& -k_p y_{a,1} - k[2 y_{a,1} - y_{a,2} - \lambda (2 y_{a,1} - y_{a,2})] \cr
	&+&\int_{t_0}^{t} dt' \Sigma_L^{+}(t-t') y_{a,1}(t') +  \eta_{a,L},
	\label{e:EOM-ant-sym-1-diff-bath-Rub}\\
	m_0 \ddot{y}_{a,N} &=& -k_p y_{a,N} - k\Big[2 y_{a,N} - y_{a,N-1} \cr 
	&-&\lambda (2 y_{a,N} - y_{a,N-1})\Big]\cr
	&+&\int_{t_0}^{t} dt' \Sigma_R^{+}(t-t') y_{a,N}(t') + \eta_{a,R}. \quad
	\label{e:EOM-ant-sym-N-diff-bath-Rub}
	\eea
We notice that, letting $t_0 \to -\infty$ and taking a Fourier transformation to the frequency domain, the above EOM in Eqs.~\ref{e:EOM-sym-1-diff-bath-Rub}, \ref{e:EOM-sym-N-diff-bath-Rub}, \ref{e:EOM-ant-sym-1-diff-bath-Rub}, and \ref{e:EOM-ant-sym-N-diff-bath-Rub} transform into Eqs.~\ref{e:fm5}, \ref{e:fm7}, \ref{e:fm6}, and \ref{e:fm8}, respectively.

\subsection{Linear response heat current}	
	
We define the heat current at the left boundary of the network as the rate of work done by the baths, which is given by the expression in Eq.~\ref{e:heat-current-J2}. It can be re-expressed in terms of symmetric and anti-symmetric variables as 
	\bea
	J_{\rm II} &=& -\Big[ \dot{y}_{s,1} \eta_{s,L} + \dot{y}_{a,1} \eta_{a,L} \cr
	&+& \left( \frac{\dot{y}_{s,1} + \dot{y}_{a,1}}{\sqrt{2}} \right) \int_{-\infty}^{t} dt' \Sigma_L^{+}(t-t') \left(\frac{y_{s,1} + y_{a,1}}{\sqrt{2}} \right)   \cr
	&+& \left( \frac{\dot{y}_{s,1} - \dot{y}_{a,1}}{\sqrt{2}} \right) \int_{-\infty}^{t} dt' \Sigma_L^{+}(t-t') \left( \frac{y_{s,1} - y_{a,1}}{\sqrt{2}} \right) \Big]. \nonumber\\
	\eea
Now going into the Fourier modes in the frequency domain, we have the noise averaged heat current:
	\bea
	\bra J_{\rm II} \ket &=& -\int_{-\infty}^{\infty} d\omega \int_{-\infty}^{\infty} d\omega' e^{-i(\omega + \omega')t} \Big \bra i\omega \, \tl{y}_{s,1}(\omega) \tl{\eta}_{s,L}(\omega') \cr
	&+&   i \omega \tl{\Sigma}_L^+(\omega') \tl{y}_{s,1}(\omega) \tl{y}_{s,1}(\omega')  + i \omega \, \tl{y}_{a,1}(\omega) \tl{\eta}_{a,L}(\omega') \cr
	  &+&    i \omega \tl{\Sigma}_L^+(\omega') \tl{y}_{a,1}(\omega) \tl{y}_{a,1}(\omega') \Big \ket.
	\eea
From Eqs.~\ref{e:fm5}, \ref{e:fm6},  \ref{e:fm7}, and \ref{e:fm8}, we notice that  
	\bea
	\tl{y}_{s,1} &=& [\bar{G}^{+}_s]_{1,1} \left(\frac{\tl{\eta}_{1,L}  + \tl{\eta}_{2,L}}{\sqrt{2}} \right) + [\bar{G}^{+}_s]_{1,N} \left(\frac{\tl{\eta}_{1,R} + \tl{\eta}_{2,R} }{\sqrt{2}} \right), \cr
	\tl{y}_{a,1} &=& [\bar{G}^{+}_a]_{1,1} \left(\frac{\tl{\eta}_{1,L} - \tl{\eta}_{2,L}}{\sqrt{2}} \right) + [\bar{G}^{+}_a]_{1,N} \left(\frac{\tl{\eta}_{1,R} - \tl{\eta}_{2,R} }{\sqrt{2}} \right). \nonumber \\
	\eea
Here, $\bar{G}^{+}_{s/a}$ are the inverses of the matrices $\tl{Z}^{s/a}$, defined as 
	\bea
	\tl{Z}^{s/a} = \begin{bmatrix}
	\tl{z}_{L \pm} & -\lambda_{\pm} & 0 & \cdots & & 0 \\
	-\lambda_{\pm}  & \tl{z}_{\pm}  & -\lambda_{\pm} & \ddots & &\vdots \\
	\vdots & \ddots & \ddots & \ddots &  \ddots &  \vdots \\
	\vdots &  & \ddots & \ddots & \ddots & \vdots\\
	\vdots  & & & -\lambda_{\pm} & \tl{z}_{\pm} 	&  -\lambda_{\pm}\\
	0 &  \cdots & \cdots & & 	-\lambda_{\pm} & \tl{z}_{R \pm}  \\
	\end{bmatrix}, \quad
	\label{e:symmetric-and-anti-symmetric-Z-matrix}
	\eea		
where, $\tl{z}_{b \pm} = -m_0 \omega^2 + k_p + 2 \lambda_{\pm} - \tilde{\Sigma}_{b}^{+}(\omega)$, $\tl{z}_{\pm} = -m_0 \omega^2 + k_p + 2 \lambda_{\pm}$ and $\lambda_{\pm} = k(1 \pm \lambda)$ for $b = L,R$. Notice that here, $\lambda_{+}$ and $\lambda_{-}$ correspond to $\tl{Z}^{s}$ and $\tl{Z}^{a}$, respectively. Now using the fluctuation-dissipation relation (see also Eq.~(\ref{e:FD-theorem}) in Appendix \ref{a:Single-bath-at-each-boundary-Rubin-Model}),
	\beq
	\bra \tl{\eta}_{a,b}(\omega) \tl{\eta}_{a',b'}(\omega') \ket = \frac{ \tl{\Gamma}_b(\omega) \hbar}{\pi} (1 + f_b) \delta_{b,b'} \delta_{a,a'}\delta(\omega + \omega'),
	\eeq
where $a, a' =1,2$, for  noise averaging to get the right-part of the heat current as
	\bea
	\bra J_{\rm II}^R \ket &=& -\int_{-\infty}^{\infty} d\omega \,\Big[ (|[\bar{G}^{+}_s(\omega)]_{1,N}|^2 + |[\bar{G}^{+}_a(\omega)]_{1,N}|^2)  \times\cr
	&& \tl{\Gamma}_L(\omega) \tl{\Gamma}_R(\omega) \frac{\hbar \omega}{\pi}(1 + f_R) \Big].
	\eea
Similarly, finding the contribution from the left baths implies the expression for the total current as in Eq.~\ref{e:heatcur-RG-J2}. Now, we choose $\tl{\Gamma}_L(\omega)=\tl{\Gamma}_R(\omega)$ (which are nonzero only when $|\omega| < 2\sqrt{k_0}$) by setting $\gamma_L^{(1)/(2)} = \gamma_R^{(1)/(2)} = \gamma$ to get 
	\bea
	\bra J_{\rm II} \ket &=& \int_{-2\sqrt{k_0}}^{2 \sqrt{k_0}} d\omega \, \Big[ (|[\bar{G}^{+}_s(\omega)]_{1,N}|^2 + |[\bar{G}^{+}_a(\omega)]_{1,N}|^2)  \cr
	&& \times \frac{m_0 \omega^2 \gamma^4}{k_0^3}\left( 1 - \frac{m_0 \omega^2}{4k_0} \right)  \frac{\hbar \omega}{\pi}(f_L - f_R) \Big].
	\label{e:quant-current-two-bath-Rubin}
	\eea
In the linear response regime, we have
	\bea
	\bra J_{\rm II} \ket &=& \frac{k_B \Delta T}{\pi} \int_{-2\sqrt{k_0}}^{2 \sqrt{k_0}} d\omega \, \Big[ \Big(|[\bar{G}^{+}_s(\omega)]_{1,N}|^2 \cr
	&& + |[\bar{G}^{+}_a(\omega)]_{1,N}|^2 \Big) \frac{m_0 \omega^2 \gamma^4}{k_0^3}\left( 1 - \frac{m_0 \omega^2}{4k_0} \right) \cr
	&& \times \left( \frac{\hbar \omega}{2 k_B T} \right)^2 {\rm csch}^2 \left( \frac{\hbar \omega}{2 k_B T} \right) \Big]. 
	\label{e:quant-current-two-bath-Rubin-linear-response}
	\eea
Next, we obtain an analytical expression for the classical current in the thermodynamic limit.

\subsection{Thermodynamic limit and an analytic expression for the classical current}	

To find an analytical expression for the classical current, we manipulate the properties of the matrices $\tl{Z}^{s/a} = \lambda_{\pm}\cdot \bar{Z}_{s/a}$ in the large $N$ limit. Given $N$ number of lattice sites in the network, $\bar{Z}_{s/a}$ matrices are $N \times N$ tridiagonal matrices of the form given in Sec. \ref{s:Two-baths}. Now, we define $\tl{a}_{\pm} = ((-m_0 \omega^2 + k_p)/\lambda_{\pm}) + 2$ and $\epsilon_{L,R} = \tl{\Sigma}_{L,R}^{+}/\lambda_{\pm}$ for $\bar{Z}_s$ and $\bar{Z}_a$, respectively. Since these are tridiagonal matrices, we obtain
	\beq
	|[\bar{G}^{+}_{s/a}]_{1,N}|^2 = \frac{1}{(\lambda_{\pm})^2 |\det [\bar{Z}_{s/a}]_N|^2}.
	\eeq	
We simplify the expression for determinant by assuming $\tl{a}_{\pm} = 2 \cos q$. This implies $m_0 \omega_{s/a}^2 = k_p + 4 \lambda_{\pm} \sin^2(q/2)$ for $\bar{Z}_s$ and $\bar{Z}_a$, respectively. As explained in Appendix \ref{a:Single-bath-at-each-boundary-Rubin-Model}, we find the determinant of the matrices $\bar{Z}_{s/a}$ in the large $N$ limit (see Eqs.~\ref{e:determinant-of-tri-diag-matrix}, \ref{e:A-and-B-functions-for-determinant}, and \ref{e:det-square-of-tri-diag-matrix}). In this case, from the matrix elements of $\bar{Z}_s$ and $\bar{Z}_a$ along with Eqs.~\ref{e:Memory-Kernel-Rubin-Fourier} and \ref{e:A-and-B-functions-for-determinant}, we get ($\lambda_{\pm}$ correspond to $\bar{Z}_s$ and $\bar{Z}_a$, respectively):
\begin{widetext}
	\bea
	2(a_1 b_2 - a_2 b_1) &=& \frac{2 \gamma^2 \sqrt{(k_p + 4 \lambda_{\pm} \sin^2(q/2))(4 k_0 - k_p - 4 \lambda_{\pm}\sin^2(q/2))}}{k_0^4}\cr
	&& \times \frac{\left(\gamma^4 + k_0^2 \lambda_{\pm}^2 - \lambda_{\pm} \gamma^2 (\cos q) (2 k_0 - k_p - 4 \lambda_{\pm} \sin^2(q/2))\right) \sin q}{\lambda_{\pm}^3}.
	\label{e:large-N-limit-symmetric-and-anti-symmetric-integrand-terms}
	\eea
\end{widetext}
Here, the classical current can be expressed as in Eq.~\ref{e:heatcur-RG-J2-clas} in the linear response regime. We use the expression for $\Delta_N$ (see Eq.~\ref{e:determinant-of-tri-diag-matrix}) in the large $N$ limit to re-express the symmetric and anti-symmetric parts of the current as
	\beq
	\bra J_{\rm II}^{cl} \ket^{s/a} = \frac{k_B \Delta T}{\pi} \int_{0}^{\pi} dq \frac{g^{s/a}_1(q)}{(1- g^{s/a}_2(q)^2)^{1/2}}.
	\eeq
Now, we use the form of Eq.~\ref{e:large-N-limit-integrand-simplification-relation} and Eq.~\ref{e:large-N-limit-symmetric-and-anti-symmetric-integrand-terms} to simplify the integrand in the expression for classical current (where we define $\Lambda_{\pm} = \gamma^4 + k_0^2 \lambda_{\pm}^2$) as:
\begin{widetext}
	\bea
	\frac{g^{s/a}_1(q)}{(1- g^{s/a}_2(q)^2)^{1/2}} = \frac{ (\sin^2 q) \gamma^2 \lambda_{\pm}^2 \sqrt{4 k_0 - k_p - 4 \lambda_{\pm} \sin^2(q/2)}}{2 \sqrt{m_0}\left( \Lambda_{\pm} -  \gamma^2 \lambda_{\pm} (\cos q) (2 k_0 - k_p - 4 \lambda_{\pm} \sin^2(q/2)) \right)}. 
	\eea
The substitution $x = \cos q$ simplifies the symmetric and anti-symmetric parts of the classical current:
	\bea
	\bra J_{\rm II}^{cl} \ket^{s/a} = \frac{k_B\Delta T}{2 \pi \sqrt{m_0}} \int_{-1}^{1}  \frac{dx \,\gamma^2 \lambda_{\pm}^2 \sqrt{1 - x^2} \sqrt{4 k_0 - k_p - 2 \lambda_{\pm}(1-x)}}{ \Lambda_{\pm} - \gamma^2 \lambda_{\pm} x (2 k_0 - k_p -2\lambda_{\pm}(1-x))}.\label{clcurr2}
	\eea
We may evaluate this integral in terms of elliptic integrals as in the previous case to obtain (for $0 \leq \lambda < 1$)
	\bea
	\bra J_{\rm II}^{cl} \ket^{s/a} &=& \frac{k_B \Delta T}{4 \pi \sqrt{m_0}\sqrt{k_p - 4 k_0 + 4 \lambda_{\pm}}} \left[ -2(k_p + 4 \lambda_{\pm} - 4 k_0 ) (E(\phi^{s/a}, \zeta^{s/a}) + E(\zeta^{s/a})) \right. \cr
	&& \left. + 4(\lambda_{\pm} - k_0) (F(\phi^{s/a}, \zeta^{s/a}) + K(\zeta^{s/a})) \right.\cr
	&&\left.+ (k_p + 2 \lambda_{\pm} -2 k_0) \left(\Pi(\xi_{-}^{s/a}, \phi^{s/a}, \zeta^{s/a}) + \Pi(\xi_{+}^{s/a}, \phi^{s/a}, \zeta^{s/a}) + \Pi(\xi_{-}^{s/a}, \zeta^{s/a}) + \Pi(\xi_{+}^{s/a}, \zeta^{s/a}) \right) \right. \cr
	&&\left. + \frac{\gamma^2  \left(4 k_0^2 - 4 k_0 (k_p + 2 \lambda_{\pm}) + k_p^2 + 4 k_p \lambda_{\pm} - 4 \lambda_{\pm}^2 \right) + 4 k_0^2 \lambda_{\pm}^2  + 4 \gamma^4}{\sqrt{8 \gamma^2 k_0^2 \lambda_{\pm}^2 + \gamma^4 (-2 k_0 + k_p + 2 \lambda_{\pm})^2 + 8 \gamma^6 }} \right. \cr
	&&\left. \times \left(\Pi(\xi_{-}^{s/a}, \phi^{s/a}, \zeta^{s/a}) - \Pi(\xi_{+}^{s/a}, \phi^{s/a}, \zeta^{s/a}) + \Pi(\xi_{-}^{s/a}, \zeta^{s/a}) - \Pi(\xi_{+}^{s/a}, \zeta^{s/a}) \right) \right].
	\label{e:analytical-formula-for-sym-antsym-currents}	
	\eea
Here, $\lambda_{\pm}$ correspond to $\bra J_{\rm II}^{cl} \ket^{s/a}$, respectively, and 
	\bea
	\phi^{s/a} = i \sinh^{-1} \left(\sqrt{\frac{4 k_0 - k_p -4 \lambda_{\pm}}{k_p - 4 k_0 }}\right), \quad
	\zeta^{s/a} = \frac{k_p - 4 k_0}{k_p + 4 \lambda_{\pm} - 4 k_0}
	\eea
and	
	\bea
	\xi_{\pm}^{s/a} =\frac{2 \gamma ^2 (k_p- 4 k_0)}{\gamma^2  (-6 k_0 + k_p + 2 \lambda_{\pm}) \pm  \sqrt{8 \gamma^2 k_0^2 \lambda_{\pm}^2 + \gamma^4 (-2 k_0  + k_p + 2 \lambda_{\pm})^2 + 8 \gamma^6}}.  
	\label{e:xi-pm-sym-and-anti-sym}
	\eea
Thus, we have an analytical expression for classical current in the case of two different baths at each boundary of the network.

\end{widetext}

\subsection{Temperature dependence of low-temperature $\bra J_{\rm II}\ket$}

Here, we analyse the temperature dependence of the quantum current $\bra J_{\rm II}\ket$, obtained when two different baths are coupled at each boundary of the system. From the above analysis, we know that the symmetric part of the quantum current can be expressed as:

\vspace{0.5cm}
\begin{widetext}
	\bea
	\bra J_{\rm II}\ket^s = \frac{\hbar^2 \Delta T}{8 \pi k_B T^2 \sqrt{m_0}} \int_{0}^{\pi}  dq \Big[\frac{ \gamma^2 \lambda_{+}^2 \sin^2(q) \sqrt{4 k_0  - m_0 \omega_s^2}}{\Lambda_{+} - \gamma^2 \lambda_{+}(\cos q)(2 k_0 - m_0 \omega_s^2)} \times \omega_s^2 \, {\rm csch}^2\left( \frac{\hbar \omega_s}{2 k_B T}\right)\Big].
	\label{e:j2-symmetric-quantum-current}
	\eea
Here, $m_0 \omega_{s}^2 = k_p + 4 \lambda_{+} \sin^2(q/2)$. In the small $q$ limit, this expression simplifies to:
	\beq
	\bra J_{\rm II}\ket^s = \frac{\hbar^2 \Delta T}{8 \pi k_B T^2 \sqrt{m_0} } \int_{0}^{\pi} dq \Big[  \frac{ \gamma^2 \lambda_{+}^2 q^2 \sqrt{4 k_0 - k_p - \lambda_{+} q^2}}{ \Lambda_{+} - \gamma^2 \lambda_{+} (1- q^2/2) (2 k_0 - k_p - \lambda_{+} q^2)} \left( \frac{k_p}{m_0} + \frac{\lambda_{+} q^2}{m_0}\right) 4 e^{-\frac{ \hbar \omega_0}{k_B T}} e^{- \frac{\lambda_{+}q^2}{2 k_B T m_0 \omega_0}} \Big]. 
	\label{e:j2-symmetric-quantum-current-small-q}
	\eeq
\end{widetext}	
Following the similar arguments as in Appendix \ref{a:Single-bath-at-each-boundary-Rubin-Model}, for $k_p \neq 0$, from Eq.~\ref{e:j2-symmetric-quantum-current-small-q} we get the leading-order $T$-dependence as:	
	\beq
	\bra J_{\rm II}\ket^s \sim \frac{e^{\frac{-\hbar \omega_0}{k_B T}}}{\sqrt{T}} \quad  \text{with} \quad \omega_0 = \sqrt{\frac{k_p}{m_0}}.
	\eeq
Now, to understand the temperature dependence of the current when $k_p = 0$, we notice that the denominator in the integrand in Eq.~\ref{e:j2-symmetric-quantum-current} for small $q$ limit takes the form: 
	\beq
	(\gamma^2 - k_0 \lambda_{+})^2 + (k_0 \lambda_{+} \gamma^2 + \gamma^2 \lambda_{+}^2) q^2.
	\eeq  
Thus, analytically, we get $\bra J_{\rm II}\ket^s \sim T$, when $\gamma^2 = k_0 \lambda_{+}$ and $\bra J_{\rm II}\ket^s \sim T^3$ when $\gamma^2 \neq  k_0 \lambda_{+}$. Similar results hold for anti-symmetric part of the quantum current $\bra J_{\rm II}\ket^a$. However, it is only defined in the regular regime, where the anti-symmetric frequencies are real. Thus, for $k_p = 0$, the total quantum current $\bra J_{\rm II} \ket \sim T$ if either $\gamma^2 = k_0 \lambda_{+}$ or $\gamma^2 = k_0 \lambda_{-}$. On the other hand, if $\gamma^2 \neq k_0 \lambda_{+} \neq k_0 \lambda_{-}$, then $\bra J_{\rm II} \ket \sim T^3$. We are able to reproduce these results numerically.

\section{Two Ohmic baths at each boundary}
\label{c:Two-baths-at-each boundary-white-noise}

In this appendix, we consider the case where two types of oscillators at each end of the network are connected to two different Ohmic baths. The Ohmic baths are modeled by white noise, implying that the noises are uncorrelated at different times, leading to Markovian dynamics. Thus, in this case, the self energy of the baths are given by the expression $\tl{\Sigma}_{L/R}^{+}(\omega) =  i \gamma^o_{L/R}\omega$ (we assume both baths have the same coupling $\gamma_L^{o^{(1)}} =\gamma_L^{o^{(2)}} = \gamma_L^{o}$). We write down the EOM for the network variables as: 
	\bea
	m_0 \ddot{y}_{\alpha,1} &=& -k_p (y_{\alpha,1}) - k\left[(2 y_{\alpha,1} - y_{\alpha,2}) \right. \cr
	&+&~~ \left. \lambda (2 y_{\beta,1} - y_{\beta,2}) \right] -\gamma_L^o \dot{y}_{\alpha,1} + \eta_{\alpha,L}  \quad \text{and} \cr
	m_0 \ddot{y}_{\alpha,N} &=& -k_p (y_{\alpha,N}) -k \left[ (2 y_{\alpha,N} - y_{\alpha,N-1}) \right. \cr
	&+&~~ \left.  \lambda ( 2 y_{\beta,N} - y_{\beta,N-1}) \right] - \gamma_R^o \dot{y}_{\alpha,N} + \eta_{\alpha,R}. \qquad
	\eea	
Here, $\alpha \neq \beta$ and $\alpha, \beta = 1,2$, and $\eta_{\alpha,L/R}$ represent the noises. Next, we rewrite these EOM in terms of symmetric and anti-symmetric variables ($y_{s/a,n} = (y_{1,n} \pm y_{2,n})/\sqrt{2}$). The symmetric variables at the boundaries satisfy the EOM:	
	\bea
	m_0 \ddot{y}_{s,1} &=& -k_p (y_{s,1}) - k \big[(2 y_{s,1} - y_{s,2}) \cr
	&+& \lambda (2 y_{s,1} - y_{s,2}) \big] - \gamma_L^o  \dot{y}_{s,1} + \eta_{s,L}, \cr
	m_0 \ddot{y}_{s,N} &=& -k_p (y_{s,N}) - k \big[ (2 y_{s,N} - y_{s,N-1}) \cr
	&+& \lambda ( 2 y_{s,N} - y_{s,N-1}) \big] - \gamma_R^o \dot{y}_{s,N} + \eta_{s,R}.  \qquad
	\label{e:EOM-symmetric-white-noise}
	\eea	
Similarly, the anti-symmetric variables at the boundaries satisfy the EOM:	
	\bea
	m_0 \ddot{y}_{a,1} &=& -k_p (y_{a,1}) - k \big[(2 y_{a,1} - y_{a,2}) \cr
	&-&\lambda (2 y_{a,1} - y_{a,2}) \big] - \gamma_L^o \dot{y}_{a,1} + \eta_{a,L}, \cr
	m_0 \ddot{y}_{a,N} &=& -k_p (y_{a,N}) - k \big[ (2 y_{a,N} - y_{a,N-1}) \cr
	&-&\lambda ( 2 y_{a,N} - y_{a,N-1}) \big] - \gamma_R^o \dot{y}_{a,N} + \eta_{a,R}. \qquad
	\label{e:EOM-anti-symmetric-white-noise}
	\eea
The remaining symmetric and anti-symmetric variables $y_{s/a,n}$ (for $n = 2, \cdots, N-1$) satisfy the EOM:
	\bea
	m_0 \ddot{y}_{s/a,n} &=& -k_p y_{s/a,n} - k \big[ \left(2 y_{s/a,n} - y_{s/a, n-1} - y_{s/a,n+1} \right) \cr
	&\pm & \lambda (2 y_{s/a,n} - y_{s/a,n+1} - y_{s/a,n-1}) \big].
	\eea
In terms of the Fourier modes in frequency domain, the EOM for the boundary variables become:
	\bea
	z_{L,+} \tl{y}_{s,1}(\omega) - k(1 + \lambda) \tl{y}_{s,2}(\omega) &=&  \tl{\eta}_{s,L}, \cr
	z_{R,+} \tl{y}_{s,N}(\omega) - k(1 + \lambda) \tl{y}_{s,N-1}(\omega) &=& \tl{\eta}_{s,R}, \cr
	z_{L,-} \tl{y}_{a,1}(\omega) - k(1 - \lambda) \tl{y}_{a,2}(\omega) &=& \tl{\eta}_{a,L}, \cr
	z_{R,-}\tl{y}_{a,N}(\omega) - k(1 - \lambda) \tl{y}_{a,N-1}(\omega) &=& \tl{\eta}_{a,R}.
	\label{e:EOM-Fourier-symmetric-anti-symmmetric-white-noise}
	\eea 
Here, $z_{L/R,+} = (-m_0 \omega^2 + k_p + 2k(1 + \lambda) - i \gamma_{L/R}^o \omega) $ and $z_{L/R,-} = (-m_0 \omega^2 + k_p + 2k(1 - \lambda) - i \gamma_{L/R}^o \omega)$. 	

\subsection{Linear response heat current}		

We define the heat current as the rate of work done by the left baths on the network, which takes the form:
	\beq
	J = - \left[\dot{y}_{1,1}(-\gamma_L^o \dot{y}_{1,1} + \eta_{1,L}) + \dot{y}_{2,1}(-\gamma_L^o \dot{y}_{2,1} + \eta_{2,L})\right].
	\eeq
In terms of symmetric and anti-symmetric variables, the current becomes
	\beq
	J = -\eta_{s,L}\dot{y}_{s,1} - \eta_{a,L} \dot{y}_{a,1} + \gamma_L^o(\dot{y}_{s,1}^2 + \dot{y}_{a,1}^2).
	\eeq
Now going to the Fourier modes in frequency domain, we express the noise averaged heat current as:
	\bea
	\bra J_{\rm II}^{\rm w} \ket &=& \int_{-\infty}^{\infty} d\omega \int_{-\infty}^{\infty} d\omega' e^{-i(\omega + \omega')t} \Big \bra i\omega \, \tl{y}_{s,1}(\omega) \tl{\eta}_{s,L}(\omega')  \cr
	&+&   \gamma_L^o i \omega \, i \omega' \tl{y}_{s,1}(\omega ) \tl{y}_{s,1}(\omega') + i \omega \, \tl{y}_{a,1}(\omega) \tl{\eta}_{a,L}(\omega')  \cr
	&+& \gamma_L^o i \omega \, i \omega' \tl{y}_{a,1}(\omega )\tl{y}_{a,1}(\omega')) \Big \ket.
	\eea 	
From the EOM in Eq.~\ref{e:EOM-Fourier-symmetric-anti-symmmetric-white-noise}, we notice that 
	\bea
	\tl{y}_{s/a,1} &=& [\tl{G}^{s/a}]_{1,1} \left(\frac{ \tl{\eta}_{1,L} \pm \tl{\eta}_{2,L} }{\sqrt{2}} \right) \cr
	&+& [\tl{G}^{s/a}]_{1,N} \left(\frac{ \tl{\eta}_{1,R} \pm \tl{\eta}_{2,R}}{\sqrt{2}} \right), \cr
	\tl{y}_{s/a,N} &=& [\tl{G}^{s/a}]_{N,1} \left(\frac{ \tl{\eta}_{1,L} \pm \tl{\eta}_{2,L}}{\sqrt{2}} \right) \cr
	&+& [\tl{G}^{s/a}]_{N,N} \left(\frac{ \tl{\eta}_{1,R}  \pm \tl{\eta}_{2,R}}{\sqrt{2}} \right). 
	\eea
Here, $\tl{G}^{s/a}$ are the inverses of the matrices $Z^{s,a}$ defined as: 
	\beq
	Z^{s/a} = \begin{bmatrix}
	z_{L,\pm}  & -\lambda_{\pm} & 0 & \cdots & & 0 \\
	-\lambda_{\pm}  & z_{\pm}  & -\lambda_{\pm} & \ddots & &\vdots \\
	\vdots & \ddots & \ddots & \ddots &  \ddots &  \vdots \\
	\vdots &  & \ddots & \ddots & \ddots & \vdots\\
	\vdots  & & & -\lambda_{\pm} & z_{\pm} 	&  -\lambda_{\pm}\\
	0 &  \cdots & \cdots & & 	-\lambda_{\pm} & z_{R,\pm}  \\
	\end{bmatrix},
	\label{e:symmetric-and-anti-symmetric-Z-matrix}
	\eeq
where, $\tl{z}_{b, \pm} = -m_0 \omega^2 + k_p + 2 \lambda_{\pm} - i \gamma^o_b \omega$ and $z_{\pm} = -m_0 \omega^2 + k_p + 2 \lambda_{\pm}$. Recall $\lambda_{\pm} = k(1 \pm \lambda)$ and they correspond to $Z^{s}$ and $Z^{a}$, respectively. The noise average can be computed using the fluctuation-dissipation relation for the Ohmic baths:
	\beq
	\bra \tl{\eta}_{a,b}(\omega) \tl{\eta}_{a',b'}(\omega') \ket = \frac{ \gamma_b^o \hbar \omega}{\pi} (1 + f_b) \delta_{b,b'} \delta_{a,a'} \delta(\omega + \omega').
	\eeq
Here, $a,a' = 1,2$ and $b = L,R$. As explained in previous appendices, finding the left and right parts of the current and combining them, we derive the expression for total current:	
	\bea
	\bra J_{\rm II}^{\rm w} \ket &=&  \int_{-\infty}^{\infty} d\omega \Big[ \, (|[\tl{G}^s(\omega)]_{1,N}|^2 + |[\tl{G}^a(\omega)]_{1,N}|^2) \times \cr
	&& \gamma_L^o \gamma_R^o \frac{\hbar \omega^3}{\pi}( f_L - f_R) \Big]  = \bra J_{\rm II}^{\rm w} \ket^s  + \bra J_{\rm II}^{\rm w} \ket^a. 
	\label{e:quantum-current-white-noise-two-bath}
	\eea
In the linear response regime, this heat current takes the form:
	\bea
	\bra J_{\rm II}^{\rm w} \ket &=&  \frac{k_B \Delta T}{\pi}\int_{-\infty}^{\infty} d\omega \Big[ \, (|[\tl{G}^s(\omega)]_{1,N}|^2 + |[\tl{G}^a(\omega)]_{1,N}|^2)  \cr
	&& \times \gamma_L^o \gamma_R^o \omega^2 \left(\frac{\hbar \omega}{2 k_B T} \right)^2 {\rm csch}^2\left( \frac{\hbar \omega}{2 k_B T}\right) \Big]. 
	\label{e:quantum-current-white-noise-two-bath-linear-esponse}
	\eea
Next, we find an analytical expression for the current in the thermodynamic limit for large temperature. 
	
\subsection{Thermodynamic limit and analytical expression for classical heat current}

As explained in previous appendices, to find the analytical expression for classical current, we need to manipulate large $N$ properties of the matrix $Z^{s/a} = \lambda_{\pm} \tl{Z}^{s/a}$. Given $N$ number of lattice sites in the network, $\tl{Z}^{s/a}$ are $N\times N$ tridiagonal matrices whose offdiagonal elements are $-1$ and diagonal elements excluding the first and last one are $(-m_0 \omega^2 + k_p)/\lambda_{+,-}) + 2$, respectively. The first and last diagonal elements of $\tl{Z}^{s}$ are $z_{L,+}/\lambda_+$ and $z_{R,+}/\lambda_+$,  respectively, and those for $\tl{Z}^{a}$ are $z_{L,-}/\lambda_-$ and $z_{R,-}/\lambda_-$,  respectively. The expression for heat current in Eq.~\ref{e:quantum-current-white-noise-two-bath} involves matrix elements $[\tl{G}^{s/a}]_{1,N}$, which is given by the inverse of $Z^{s/a}$: $\tl{G}^{s/a} = (Z^{s/a})^{-1} = (\tl{Z}^{s/a})^{-1}/ \lambda_{\pm}$. Here, $\lambda_{\pm}$ appear in the expressions for $\tl{G}^{s}$ and $\tl{G}^{a}$ , respectively. Thus, we have
	\beq
	|\tl{G}^{s/a}_{1,N}| = \frac{1}{\lambda_{\pm} |\det \tl{Z}^{s/a}|}.
	\eeq
We simplify the determinant of the tridiagonal matrices $\tl{Z}^{s/a}$ in the large $N$ limit to get an analytical expression. Choosing $\gamma_L^o = \gamma_R^o = \gamma^o$, we find the expression for classical current as
	\bea
	\bra J_{\rm II}^{{\rm w}^{cl}} \ket  &=& \frac{k_B \Delta T}{\pi} \int_{-\infty}^{\infty} d\omega (\gamma^o)^2 \omega^2  (|G^s_{1N}|^2 + |G^a_{1N}|^2) \cr
	&=& \bra J_{\rm II}^{{\rm w}^{cl}}\ket^s + \bra J_{\rm II}^{{\rm w}^{cl}}\ket^a.
	\eea
To follow a similar derivation as done in Appendix \ref{b:Two-baths-at-each-boundary-Rubin-Model}, we define $\tl{a}_{\pm} = ((-m_0 \omega^2 + k_p)/\lambda_{\pm}) + 2$ and $\epsilon_{L,R} = i \omega \gamma^o_{L/R}/\lambda_{\pm}$ for $\tl{Z}_s$ and $\tl{Z}_a$, respectively. Now with the substitution $\tl{a}_{\pm} = 2 \cos q$, in the large $N$ limit (see Eq.~\ref{e:large-N-limit-identity}), we re-express the symmetric and anti-symmetric parts of the classical current in the form:
	\beq
	\bra J_{\rm II}^{{\rm w}^{cl}} \ket^{s/a} = \frac{k_B \Delta T}{\pi} \int_{0}^{\pi} dq \frac{g^{s/a}_1(q)}{(1- g^{s/a}_2(q)^2)^{1/2}},
	\eeq
where, the integrand is given explicitly as: 
	\beq
	\frac{g^{s/a}_1(q)}{(1- g^{s/a}_2(q)^2)^{1/2}} =  \frac{\gamma^o \lambda_{\pm}^2}{m_0} \frac{\sin^2 (q)}{\Lambda_{s/a} - \zeta_{s/a} \cos(q)}.
	\eeq
Here,
	\bea
	\Lambda_{s/a} &=& \lambda_{\pm}^2 + \frac{(k_p + 2 \lambda_{\pm}) (\gamma^o)^2}{m_0} \quad \text{and} \cr
	\zeta_{s/a} &=& \frac{2 \lambda_{\pm} (\gamma^o)^2}{m_0}.
	\eea		
Thus, the expression for the classical current is 
	\bea
	\bra J_{\rm II}^{{\rm w}^{cl}} \ket &=& \bra (J_{\rm II}^{\rm w})^{cl}\ket^s + \bra (J_{\rm II}^{\rm w})^{cl}\ket^a \cr
	&=& \frac{\gamma^o k_B \Delta T}{m_0} \Big( \frac{\lambda_{+}^2 }{\zeta_s^2} (\Lambda_s - \sqrt{\Lambda_s^2 - \zeta_s^2}) \cr
	&+& \frac{\lambda_{-}^2 }{\zeta_a^2} (\Lambda_a - \sqrt{\Lambda_a^2 - \zeta_a^2}) \Big).
	\eea

\subsection{Temperature dependence of low-temperature $\bra J_{\rm II}^{\rm w} \ket$}

Now, we analyse the temperature dependence of the low-temperature quantum heat current in Eq.~\ref{e:quantum-current-white-noise-two-bath}. In order to understand the role of pinning strength on the temperature dependence, we analyse two different cases, one  with $k_p \neq 0$ and the other with $k_p = 0$. 

We notice that for $N \to \infty$, in the linear response regime, the symmetric part of the quantum current is
	\beq
	\bra J_{\rm II}^{\rm w} \ket^s = \frac{\gamma^o \lambda_{+}^2 \hbar^2 \Delta T}{4 \pi m_0 k_B T^2} \int_{0}^{\pi} \frac{ dq \, \sin^2 q \, \omega^2}{\Lambda_s - \zeta_s \cos q}  \: {\rm csch}^2\left(\frac{\hbar \omega}{2 k_B T} \right).
	\eeq
Here, $\omega^2 = (k_p + 4 \lambda_{+} \sin^2(q/2))/m_0$. As explained in the previous appendices, by taking the small $q$ limit to understand the temperature dependence of the quantum current for $k_p \neq 0$, we find 
	\bea
	\bra J_{\rm II}^{\rm w} \ket^s &\approx& \frac{\gamma^o \lambda_{+}^2 \hbar^2 \Delta T}{4 \pi m_0 k_B T^2} \int_{0}^{\pi} \Big[  \frac{dq \, q^2}{\Lambda_s - \zeta_s (1- q^2/2)} \cr
	&&\times \left(\frac{k_p}{m_0} + \frac{\lambda_{+}q^2}{m_0} \right) 4 e^{- \frac{\hbar \omega_0}{k_B T}}e^{-\frac{\hbar \lambda_{+} q^2}{2 k_B T m_0 \omega_0}} \Big], \qquad
	\eea
where $\omega_0 = \sqrt{k_p/ m_0}$. By substituting $x = (\hbar \lambda_{+} q^2)/(2 m_0 \omega_0 k_B T)$, we obtain the leading-order $T$-dependence: $\bra J_{\rm II}^{\rm w} \ket^s \sim (e^{-\hbar \omega_0/(k_B T)})/\sqrt{T}$.

For $k_p =0$,  we have $\Lambda_s = \lambda_{+}^2 + \zeta_s$ and $\omega^2 \approx \lambda_{+}q^2/m_0$ in the small $q$ limit. Thus, we find  
	\bea
	\bra J_{\rm II}^{\rm w} \ket^s &\approx& \frac{\gamma^o \lambda_{+}^2 \hbar^2\Delta T}{4 \pi m_0 k_B T^2}\int_{0}^{\pi} dq \Big[\frac{q^2}{\lambda_{+}^2 + \zeta_s (q^2/2) } \cr
	 && \times \frac{\lambda_{+} q^2}{m_0} \: {\rm csch}^2\left(\frac{\hbar \omega}{2 k_B T} \right) \Big].
	\eea
By substituting $x = \hbar \omega/(2 k_B T)$, and  noticing that $\omega \propto q$, we get the leading order $T$-dependence $\bra J_{\rm II}^{\rm w} \ket^s \sim T^3$. These observations agree with the results already known from the heat transport of ordered harmonic lattices \cite{Roy_2008}. Similar temperature dependence holds for the anti-symmetric part of  the current $\bra J_{\rm II}^{\rm w} \ket^a$ as well. However, this current is bounded only in the regular regime.

\subsection{Comparison of classical heat current for Rubin bath and Ohmic bath}
	
	\begin{figure}[h!]
	\includegraphics[width=\linewidth, height=0.7\linewidth]{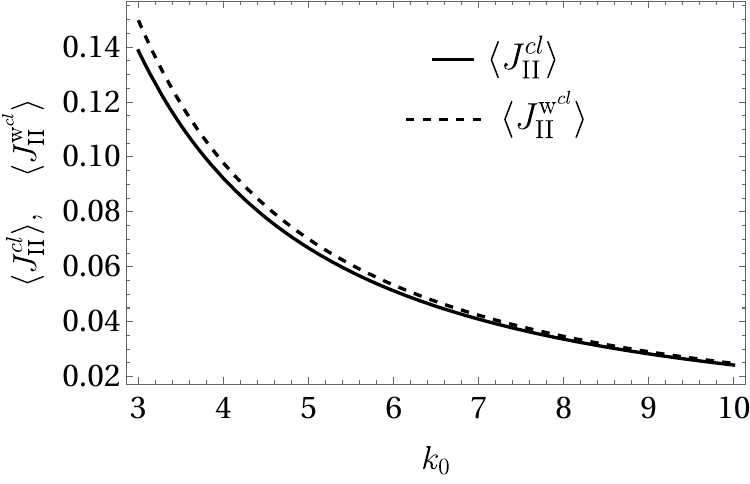}
	\caption{Comparison between the classical heat currents $\bra J_{\rm II}^{cl} \ket$ and $\bra J_{\rm II}^{{\rm w}^{cl}} \ket$ as a function of the bath spring constant $k_0$. The $y$-axis represents the heat currents multiplied by the common factor $\pi/ (k_B \Delta T)$. We set the parameters $\lambda = 0.7, \gamma = 0.5 k, k_p = 0$ and  $m_0 =  k = \hbar = k_B = 1$ in the numerical analysis. }
	\label{f:comparison-classical-J2-J2w}
	\end{figure}

We compare the classical heat currents obtained for heat baths with different spectral properties. To compare the analytical expressions for classical currents obtained through Rubin baths to that of Ohmic baths, we notice that the coupling strengths $\gamma$ and $\gamma^o$ should satisfy the condition:
	\beq
	\gamma^o = \frac{\sqrt{m_0} \gamma^2}{k_0^{3/2}}.
	\label{e:relation-between-coupling-Rubin-and-Ohmic}
	\eeq
This identity is obtained by comparing the self energies $\tl{\Sigma}^{+}(\omega)$ of both types of baths at small frequencies. We notice that the values of current become almost equal as we increase the value of $k_0$ as shown in Fig.~\ref{f:comparison-classical-J2-J2w}. This behavior is expected, as Rubin baths with higher bandwidths exhibit properties resembling Ohmic baths (See \cite{Avijit2020} for a detailed analysis).

\bibliography{bibliographyCSF}

\end{document}